\documentclass[aps,pra, twocolumn, secnumarabic,amsmath,amssymb,showpacs,superscriptaddress,floatfix]{revtex4}
\usepackage{placeins}
\usepackage{graphicx}
\usepackage{bm}
\usepackage{epsfig}
\usepackage{subfigure}
\newbox\subfigbox
\makeatletter
\newenvironment{subfloat}%
{\def\caption##1{\gdef\subcapsave{\relax##1}}%
  \let\subcapsave\@empty%
  \setbox\subfigbox\hbox%
  \bgroup}%
{\egroup%
  \subfigure[\subcapsave]{\box\subfigbox}}%
\makeatother

\def\<{\begin{equation}}
\def\>{\end{equation}}
\batchmode

\begin{document}

\title{ Fractional Quantum Hall Effect in Optical Lattices}
\author{M. Hafezi}
\email{hafezi@fas.harvard.edu} \affiliation{Physics Department,
    Harvard University, Cambridge, MA - 02138}
\author{A. S. S\o rensen}
\affiliation{QUANTOP, Danish National Research Foundation Centre of
Quantum Optics, Niels Bohr Institute,
 University of Copenhagen, DK-2100 Copenhagen \O,
Denmark.}
\author{E. Demler}
\affiliation{Physics Department,
    Harvard University, Cambridge, MA - 02138}
\author{M. D. Lukin}

\affiliation{Physics Department,
    Harvard University, Cambridge, MA - 02138}

\begin{abstract}

We analyze a recently proposed method to create fractional quantum
Hall (FQH) states of  atoms confined in optical lattices [A.
S{\o}rensen {\it et al.}, Phys. Rev. Lett. {\bf 94} 086803 (2005)].
Extending the previous work, we investigate conditions under which
the FQH effect can be achieved for  bosons on a lattice with an
effective magnetic field and finite onsite interaction. Furthermore,
we characterize the ground state in such systems  by calculating
Chern numbers which can provide direct signatures of  topological
order and explore regimes where the characterization in terms of
wavefunction overlap fails. We also discuss various issues which are
relevant for the practical realization of such FQH states with ultra
cold atoms in an optical lattice, including the presence of the
long-range dipole interaction which can improve the energy gap and
stabilize the ground state. We also investigate a new detection
technique based on Bragg spectroscopy to probe these system in an
experimental realization.
\end{abstract}


\maketitle

\title{Bosonic system on a lattice with a Magnetic Hamiltonian}

\date{\today}

\section{Introduction}
With recent advances in the field of ultra-cold atomic gases,
trapped Bose-Einstein condensates (BEC's) have become an important
system to study many-body physics such as quantum phase transitions.
In particular, the ability to dynamically control the lattice
structure and the strength of interaction as well as the absence of
impurities in BEC's confined in optical lattices, have led to the
recent observation of the superfluid to Mott-insulator transition
\cite{greiner, Mandel03, Campbell06, Winkler06, Folling05}. At the
same time, there has been a tremendous interest in studying rotating
BEC's in harmonic traps; at sufficient rotation an Abrikosov lattice
of quantized vortices has been observed \cite{aboshaeer} and the
realization of strongly correlated quantum states similar to the
fractional quantum Hall states has been predicted to occur at higher
rotation rates \cite{wilkin2000,cooper2001,cooper2005}. In these proposals,
the rotation can play the role of an effective magnetic field for
the neutral atoms and in analogy with electrons, the atoms may enter
into a state described by the Laughlin wavefunction, which was
introduced to describe the fractional quantum Hall effect. While
this approach yields a stable ground state separated from all
excited states by an energy gap, in practice this gap is rather small because of
the weak interactions among the particles in the magnetic traps
typically used. In optical lattices, the interaction energies are
much larger because the atoms are confined in a much smaller volume,
and the realization of the fractional quantum Hall effect in optical
lattices could therefore lead to a much higher energy gap and be much more robust. In a recent paper \cite{sorensen}, it was shown that
it is indeed possible  to realize the fractional quantum Hall effect
in an optical lattice, and that the energy gap achieved in this
situation is a fraction of the tunneling energy, which can be
considerably larger than the typical energy scales in a magnetic
trap.

In addition to being an interesting system in its own right,
the fractional quantum Hall effect, is also interesting from the
point of view of topological quantum computation \cite{kitaev}. In
these schemes quantum states with fractional statistics can
potentially perform fault tolerant quantum computation. So far,
there has been no direct experimental observation of fractional
statistics although some signatures has been observed in electron
interferometer experiments \cite{goldman1,goldman2}. Strongly
correlated quantum gases can be a good alternative where the systems
are more controllable and impurities are not present. Therefore,
realization of fractional quantum Hall states in atomic gases can be
a promising resource for topological quantum computation in the
future.

As noted above the FQH effect can be realized by simply rotating and
cooling atoms confined in a harmonic trap. In this situation, it can
be shown that the Laughlin wavefunction exactly describes the ground
state of the many body system \cite{wilkin2000, paredes01}. In an optical lattices, on the other
hand, there are a number of questions which need to be addressed.
First of all, it is unclear to which extent the lattices modifies
the fractional quantum Hall physics. For a single particle, the
lattice modifies the energy levels from being simple Landau levels
into  the fractal structure known as the Hofstadter butterfly
\cite{hofstadter}. In the regime where the magnetic flux going
through each lattice $\alpha$ is small, one expects that this will
be of minor importance and in Ref.~\onlinecite{sorensen} it was
argued that the fractional quantum Hall physics persists until
$\alpha\lesssim 0.3$. In this paper, we extent and quantify predictions carried out
in Ref.~\onlinecite{sorensen} . Whereas Ref.~\onlinecite{sorensen}
only considered the effect of infinite onsite interaction, we extent
the analysis to finite interactions. Furthermore, where
Ref.~\onlinecite{sorensen} mainly argued that the ground state of
the atoms was in a quantum Hall state by considering the overlap of
the ground state found by numerical diagonalization with the
Laughlin wavefunction, we provide further evidence for this claim by
characterizing the topological order of the system by calculating
Chern numbers. These calculations thus characterize the order in the
system even for parameter regimes where  the overlap with the
Laughlin wavefunction is decreased by the lattice structure.

In addition to considering these fundamental features of the FQH states on a
lattice, which are applicable regardless of the system being used to
realize the effect, we also study a number of questions which are of
particular interest to experimental efforts towards realizing the
effect with atoms in optical lattices. In particular,  we show that adding dipole interactions between the
atoms can be used to increase the energy gap and stabilize the ground state. Furthermore, we study Bragg spectroscopy of atoms in the lattice and show that this is a viable method to identify the quantum Hall states created in an experiment, and we discuss a new
method to generate an effective magnetic field for the neutral atoms
in the lattice.

The paper is organized as follows: In
Sec.~\ref{section-no_hardcore}, we study the system with finite
onsite interaction. In Sec.~\ref{section-chern} we introduce Chern
numbers to characterize the topological order of the system. The
effect of the dipole-dipole interaction has been elaborated in
Sec.~\ref{section-dipole}. Sec.~\ref{nu1/4} studies the case of
$\nu=1/4$ filling factor. Sec.~\ref{bragg} is dedicated to explore
Bragg spectroscopy of the system. Sec.~\ref{magnetic} outlines a
new approach for generating the type of the Hamiltonian studied in this
paper.

\section{Quantum Hall state of bosons on a lattice} \label{section-no_hardcore}

\subsection{The Model}
The fractional quantum Hall effect occurs for electrons confined in
a two dimensional plane under the presences of a perpendicular
strong magnetic field. If $N$ is the number of electrons in the
system and $N_{\phi}$ is the number of magnetic fluxes measured in
units of the quantum magnetic flux $\Phi_0=h/e$, then depending on the
filling factor $\nu=N/N_{\phi}$ the ground state of the system can
form highly entangled states exhibiting a rich  behaviors, such as
incompressibility, charge density waves, and anyonic excitations
with fractional statistics. In particular, when $\nu=1/m$, where $m$ is
an integer, the ground state of the system is an incompressible
quantum liquid which is protected by an energy gap from all other states,
and in the Landau gauge is well described by the Laughlin
wavefunction \cite{laughlin}:

\<\Psi(z_1,z_2,...,z_N)=\prod_{j>k}^N (z_j-z_k)^m \prod_{j=1}^N e^{-
y^2_i/2} \label{laughlin},\>
where the integer $m$ should be odd in order to meet the antisymmetrization
requirement for fermions.

Although the fractional quantum Hall
effect occurs for fermions (electrons), bosonic systems with
repulsive interactions can exhibit similar behaviors. In particular,
the Laughlin states with even $m$ correspond to bosons. In this
article, we study bosons since the experimental implementation are
more advanced for the ultra-cold bosonic systems. We study a system
of atoms confined in a 2D lattice which can be described by the
Bose-Hubbard model \cite{jaksch98} with the Peierls substitution \cite{hofstadter,peierls},
\begin{eqnarray} H&=&-J \sum_{x,y} \hat{a}^\dag_{x+1,y} \hat{a}_{x,y}e^{-i \pi
\alpha y}+ \hat{a}^\dag_{x,y+1} \hat{a}_{x,y}e^{i \pi \alpha x}+h.c. \nonumber\\
&+& U \sum_{x,y}  \hat{n}_{x,y}( \hat{n}_{x,y}-1),
\label{eq:Hamiltonian}
\end{eqnarray}
 \\
where $J$ is the hopping energy between two neighboring sites, $U$
is the onsite interaction energy, and $2 \pi \alpha$ is the phase
acquired by a particle going around a plaquette. This Hamiltonian is
equivalent to the Hamiltonian of a U(1) gauge field (transverse
magnetic field) on a square lattice. More precisely, the
non-interacting part can be written as \< -J\sum_{<ij>}a^\dagger_i
a_j \exp\left(\frac{2\pi
i}{\Phi_0}\int^j_i \vec{A}\cdot\vec{dl}\right)\>%
where $\vec{A}$ is the vector potential for a uniform magnetic field
and the path of the integral is chosen to be a straight line
between two neighboring sites. In the symmetric gauge, the vector potential
is written as $\vec{A}=\frac{B}{2}(-y,x,0)$. Hence, $\alpha$ will be
the amount of magnetic flux going through one plaquette.

While the Hamiltonian in Eq. (\ref{eq:Hamiltonian}) occurs naturally
for charged particles in a magnetic field, the realization of a
similar Hamiltonian for neutral particles is not straightforward. As
we discuss in Sec.\ \ref{magnetic}  this may be achieved in a
rotating harmonic trap, and this have been very successfully used in
a number of experiments in magnetic traps \cite{dalibard_04,
cornell_04}, but the situation is more complicated for an optical
lattice. However,  there has been a number of proposals for lattice
realization of a magnetic field  \cite{sorensen, jaksch,mueller}, and
recently it has been realized experimentally \cite{cornell_06}. Popp {\it et al.} \cite{paredes04} have studied the realization of fractional Hall states for a few particles in individual lattice sites. A new approach for rotating the entire optical lattice is discussed in
Sec.~\ref{magnetic}. The essence of the above Hamiltonian is a
non-zero phase that a particle should acquire when it goes around a
plaquette. This phase can be obtained for example by alternating the
tunneling and adding a quadratic superlattice potential
\cite{sorensen} or by simply rotating the lattice (Sec.\ref{magnetic}). The advantage of confining ultra-cold gases in an optical lattice
is to enhance the interaction between atoms which consequently
result in a higher energy gap comparing to harmonic trap proposals
(e.g. Ref. \onlinecite{wilkin2000}). This enhancement in the energy
gap of the excitation spectrum can alleviate some of the challenges for
experimental realization of the quantum Hall state for ultra-cold
atoms.

\subsection{Energy spectrum and overlap calculations }

In order to approximate a large system, we study the
system with periodic boundary condition, i.e. on  a torus, where the
topological properties of the system is best manifested.

There are two energy scales for the system: the first is the
magnetic tunneling term, $J\alpha$, which is related to the
cyclotron energy in the continuum limit $\hbar \omega_c= 4 \pi J
\alpha$ and the second is the onsite interaction energy $U$.
Experimentally, the ratio between these two energy scales can be varied by varying
the lattice potential height \cite{greiner, jaksch98} or by  Feshbach resonances \cite{Donley02, Durr04, Winkler06}. Let us
first assume that we are in the continuum limit where  $\alpha \ll 1
$, i.e. the flux through each plaquette is a small fraction of a
flux quantum.  A determining factor for describing the system is the
filling factor $\nu=N/N_{\Phi}$, and in this study we mainly focus on the
case of $\nu=1/2$, since this will be the most experimentally
accessible regime.

We restrict ourself to the simplest boundary conditions for the single
particle states $t_s(\vec L) \psi (x_s,y_s) =\psi (x_s,y_s)$, where
$t_s(\vec{L})$ is the magnetic translation operator which translates
the single particle states $\psi(x_s,y_s)$  around the torus. The definition and detailed discussion of the boundary conditions will be elaborated in Section \ref{section-chern}. The discussed quantities in this section, such as energy spectrum, gap
and overlap, do not depend on the boundary condition angles (this is also verified by our numerical calculation).

In the continuum case, for the filling fraction $\nu=1/2$, the
Laughlin state in Landau gauge $(\vec{A}=-B y\hat{x})$ is given by
Eq. (\ref{laughlin}) with $m=2$. The generalization of the Laughlin
wavefunction to a torus takes the form \cite{read96} \<
\Psi(z_1,z_2,...,z_N)=f_{rel}(z_1,z_2,...,z_N)F_{cm}(Z)e^{-\sum_i
y^2_i/2} \label{laughlin-torus},\>\\
where $f_{rel}$ is the relative part of the wave function and is
invariant under collective shifts of all $z_i$'s by the same amount,
and $F_{cm}(Z)$ is related to the motion of the center of mass and
is only a function of  $Z=\sum_i z_i$.  For a system on a torus of the size
$(L_x \times L_y)$, we write the wavefunction with the help of theta
functions, which are the proper oscillatory periodic functions and
are defined as $\vartheta \tiny{\left[
\begin{array} {c}
a\\
b\\
\end{array} \right]}
(z|\tau)=\sum_n e^{i \pi \tau (n+a)^2 + 2\pi i (n+a)(z+b)}$ where
the sum is over all integers. For the relative part we have, \< f_{rel}=\prod_{i<j}
\vartheta \left[
\begin{array} {c}
\frac{1}{2}\\
\frac{1}{2}\\
\end{array} \right] \left(\frac{z_i-z_j}{L_x}|i \frac{L_y}{L_x} \right)^2.\>

According to a general symmetry argument by Haldane
\cite{haldane85}, the center of mass wave function $F_{cm}(Z)$ is
two-fold degenerate for the case  of $\nu=1/2$, and
 is given by
\<F_{cm}(Z)=\vartheta\left[
\begin{array} {c}
l/2+(N_{\phi}-2)/4\\
-(N_{\phi}-2)/2\\
\end{array} \right] \left(\frac{2 \sum_i z_i}{L_x}|2i
\frac{L_y}{L_x} \right)\>
where $l=0,1$ refers to the two degenerate
ground states. This degeneracy in the continuum limit is due to the
translational symmetry of the ground state on the torus, and the
same argument can be applied to a lattice system when the magnetic
length is much larger than the lattice spacing $\alpha \ll 1$. For
higher magnetic filed, the lattice structure becomes more
pronounced. However, in our numerical calculation for a moderate
magnetic field $\alpha \lesssim 0.4$, we observe a two-fold
degeneracy ground state well separated from the excited state by an
energy gap. We return to the discussion of the ground state
degeneracy in Sec.\ \ref{section-chern}.

In the continuum limit $\alpha \ll 1$, the Laughlin wavefunction is
the exact ground state of the many body system with a short range interaction \cite{haldane85b, wilkin2000, paredes01}. The reason is that
the ground state is composed entirely of states in the lowest Landau
level which minimizes the magnetic part of the Hamiltonian, the first
term in Eq.\ (\ref{eq:Hamiltonian}). The expectation value of the
interacting part of the Hamiltonian, i.e. the second term in
Eq.\ (\ref{eq:Hamiltonian}), for the Laughlin state is zero regardless
of the strength of the interaction, since it vanishes when the
particles are at the same position.

To study the system with a non-vanishing $\alpha$, we have performed
a direct numerical diagonalization of the Hamiltonian for a small
number of particles. Since we are dealing with identical particles,
the states in the Hilbert space can be labeled by specifying  the
number of particles at each of the lattice sites. In the hard-core
limit, only one particle is permitted on each lattice site,
therefore for $N$ particles on a lattice with the number of sites
equal to $(N_x=L_x/a, N_y=L_y/a)$, where $a$ is the unit lattice
side, the Hilbert space size is given by the combination $ \tiny
{\left( \begin{array}{c}
N_x N_y   \\
N \\  \end{array} \right)= \frac{N_xN_y! }{N! (N_xN_y-N)!}} $. On the other hand, in case of finite
onsite interaction,  the particles can be on top of each other, so
the Hilbert space is bigger and is given by the combination $
\tiny{ \left(
\begin{array}{c}
N+N_x N_y-1   \\
N \\   \end{array} \right)} $. In our simulations the dimension of
the Hilbert can be raised up to $ \sim 4 \cdot 10^6$ and the
Hamiltonian is constructed in the configuration space by taking into
account the tunneling and interacting terms. The tunneling term is
written in the symmetric gauge, and we make sure that the phase
acquired around a plaquette is equal to $ 2 \pi \alpha$, and that
the generalized magnetic boundary condition is satisfied when the
particles tunnels over the edge of the lattice [to be discussed in
Sec.\ \ref{section-chern}, c.f. Eq. (\ref{twist_angles})].
By diagonalizing the Hamilton, we find the two-fold degenerate ground state energy which is separated by an energy gap from the excited states and the corresponding wavefunction in the configuration space. The Lauhglin wave function
(\ref{laughlin-torus}) can also be written in the configuration
space by simply evaluating the Laughlin wave function at discrete
points,  and therefore we can compared the overlap of these two dimensional subspaces.

\begin{figure}[t]
  \centering%
  \begin{subfloat}%
\includegraphics[width=.45 \textwidth]{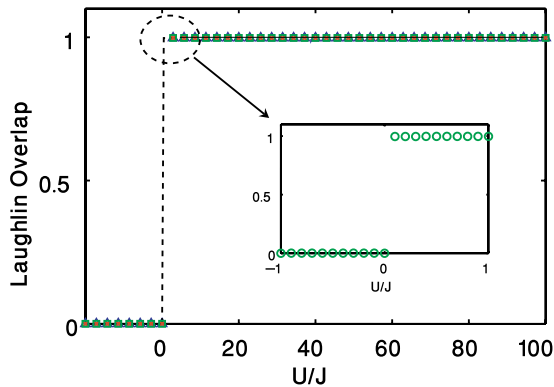}
\caption{$$ }
 \end{subfloat}%
  \hspace{0cm}%
  \begin{subfloat}%

\includegraphics[width=.45 \textwidth]{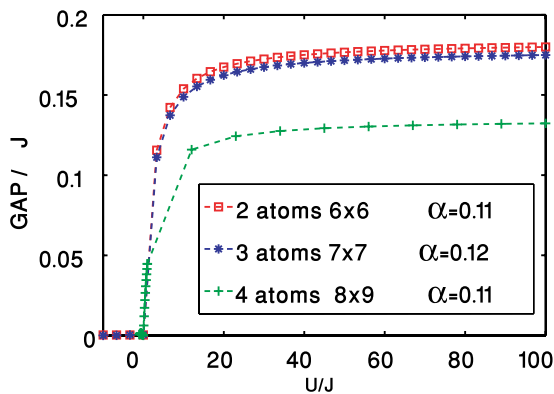}
\caption{$$ }
   \end{subfloat}%

\caption{ (color online) (a) The overlap of the ground state with the Laughlin
wavefunction. For small $\alpha$ the Laughlin wavefunction is a good
description  of the ground state for positive interaction strengths.
The inset shows the same result of small $U$. (b) The energy gap for
$ N/N_{\phi}=1/2$ as a function of interaction $U/J$ from attractive
to repulsive. For a fixed $\alpha $, the behavior does not depend on
the number of atoms. The inset define the particle numbers, lattice
sizes, and symbols  for both parts (a) and (b).}
  \label{tuning_U}
\end{figure}

\subsection{Results with the finite onsite interaction}

The energy gap above the ground state and the ground state overlap with the Laughlin
wavefunction for the case of $\nu=1/2$ in a dilute
lattice $\alpha \lesssim 0.2 $, are depicted in
figure.~\ref{tuning_U}. The Laughlin wavefunction remains a good
description of the ground state even if the strength of the
repulsive interaction tends to
zero (Fig.~\ref{tuning_U} a). Below, we discuss different limits:%

\begin{figure}
  \centering%
\begin{subfloat}%
\includegraphics[width=.45 \textwidth]{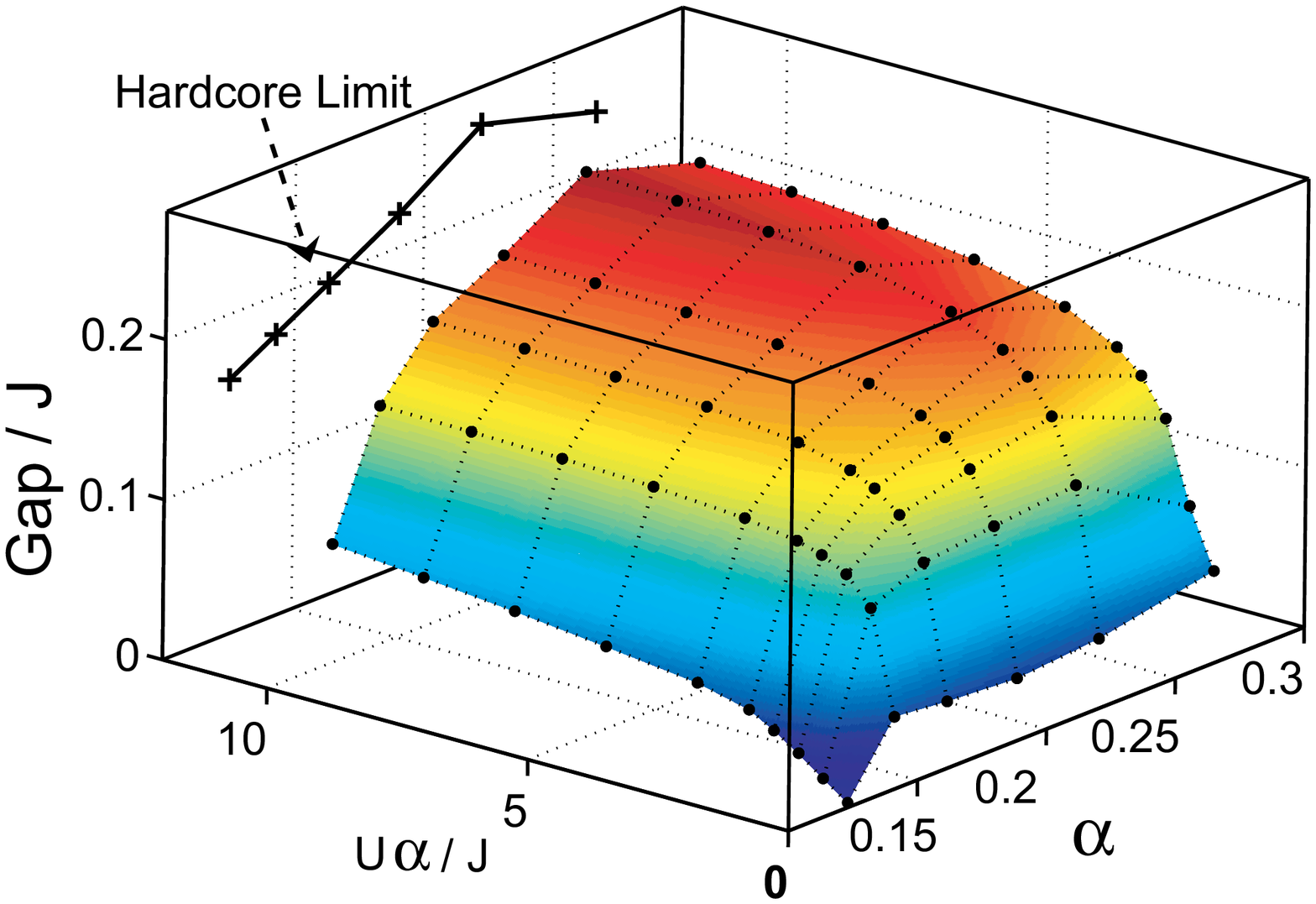}
\caption{ $$ }
 \end{subfloat}%
   \hspace{0cm}%
     \begin{subfloat}%
\includegraphics[width=.45 \textwidth]{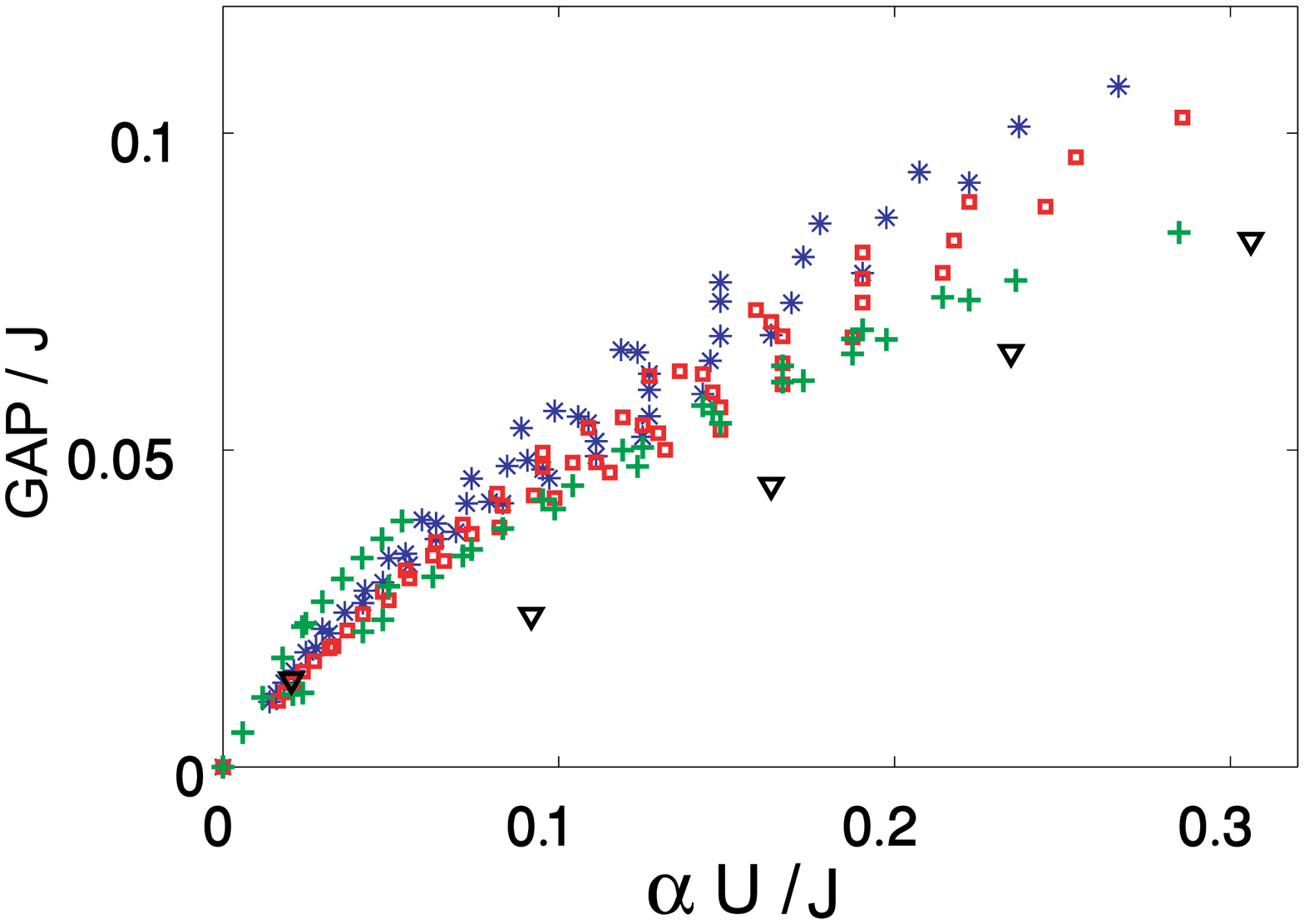}
\caption{ $$ }
 \end{subfloat}%
 \caption{(color online) (a) The energy gap as a function of $\alpha U$ and $\alpha$ for a fixed
number of atoms (N=4). The gap is calculated for the parameters
marked with dots and the surface is an extrapolation between the
points. (b) Linear scaling of the energy gap with $\alpha U$ for $U
\ll J, \alpha \lesssim 0.2$. The results are shown for $N = 2
(\square)$, $N = 3 (\ast)$, $N = 4 (+)$ and $N=5(\triangledown)$. The gap disappears for
non-interaction system, and  increases with increasing interaction
strength ($ \propto \alpha U$) and  eventually saturate to the value
in the hardcore limit.}
  \label{fig:gap}
\end{figure}

First, we consider $~U>0~,~U\gg J\alpha$: If the interaction energy
scale $U$ is much larger than the magnetic one $ (J \alpha)$, all
low energy states lie in a manifold, where the highest occupation
number for each site is one, i.e. this corresponds to the hard-core
limit. The ground state is the Laughlin state and the excited states
are various mixtures of Landau states. The ground state is two-fold
degenerate \cite{haldane85} and the gap reaches the value in for the
hard-core limit at large $U\gtrsim 3J/\alpha$, as shown in
Fig.~\ref{fig:gap} (a). In this limit the gap only depends on the
tunneling $J$ and flux $\alpha$, and the gap is a fraction of $J$.
These results are  consistent with the previous work in Ref.
\onlinecite{sorensen}.%

Secondly, we consider $~|U| \ll J\alpha$. In this regime, the
magnetic energy scale $(J \alpha)$ is much larger than the
interaction energy scale $U$. For repulsive regime ($U>0$), the
ground state is the Laughlin state and the gap increases linearly
with $\alpha U$, as shown in Fig.~\ref{fig:gap} b.

Thirdly, we study $~U=0$ where the interaction is absent and the
ground state becomes highly degenerate. For a single particle on a
lattice, the spectrum is the famous Hostadfer's butterfly
\cite{hofstadter}, while in the continuum limit  $\alpha \ll 1 $,
the ground state is the lowest Landau level (LLL). The single
particle degeneracy of the LLL is the number of fluxes going through
the surface, $N_\phi$. So in the case of $N$ bosons, the lowest
energy is obtained by putting $N$ bosons in $N_\phi$ levels.
Therefore, the many-body ground state's degeneracy should be:
$\tiny{\left(
\begin{array} {c}
N+N_\phi -1 \\
N_\phi -1 \\
\end{array} \right).}$
For example,  3 bosons and 6 fluxes gives a 336-fold degeneracy in
the non-interacting ground state.

If we increase the amount of phase (flux) per plaquette ($\alpha $)
we are no longer in the continuum limit. The Landau level degeneracy will
be replace by $\frac{L_1 L_2}{s}$ where $\alpha=r/s$ is the amount
of flux per plaquette and $r$ and $s$ are coprime \cite{fradkin}.
Then, the many-body degeneracy will be: $\tiny{\left(
\begin{array} {c}
N+\frac{L_1 L_2}{s} -1 \\
\frac{L_1 L_2}{s}-1 \\
\end{array} \right)}$.

Fourthly, we consider $~U<0~,~U\gg J\alpha$: when U is negative
(i.e. attractive interaction) in the limit strong interaction
regime, the ground state of the system will become a \textit{pillar}
state. In a pillar state, all bosons in the system condensate into a
single site. Therefore, the degeneracy of the ground state is $N_x
\times N_y $ and the ground state manifold can be spanned as,
\<  \bigoplus_i  \frac{1}{\sqrt{N!}}(a^\dagger)_i^N | vac \rangle . \>
These states will very fragile and susceptible to collapse \cite{hulet}.

In a lattice, it is also possible to  realize the fractional quantum
Hall states for attractive interaction in the limit when $ |U| \gg
J\alpha$. Assume that the occupation number of each site is either
zero or one. Since the attraction energy $U$ is very high and there
is no channel into which the system can dissipate its energy, the
probability for a boson to hop to a site where there is already a
boson is infinitesimally small. Therefore, the high energy
attraction will induce an effective hard-core constraint in the case
of ultra-cold system. The energy of these state should be exactly
equal to their hard-core ground state counterparts, since the
interaction expectation value of the interaction energy is zero for the Laughlin state. The
numerical simulation shows that these two degenerate states indeed
have a good overlap with the Laughlin wavefunction similar to their
repulsive hard-core counterparts and also their energies are equal
to the hard-core ground state. These states are very similar to
repulsively bound atom pairs in an optical lattice which have recently been experimentally observed \cite{Winkler06}.

 \begin{figure}[b]

 \begin{subfloat}
\includegraphics[width=.45\textwidth]{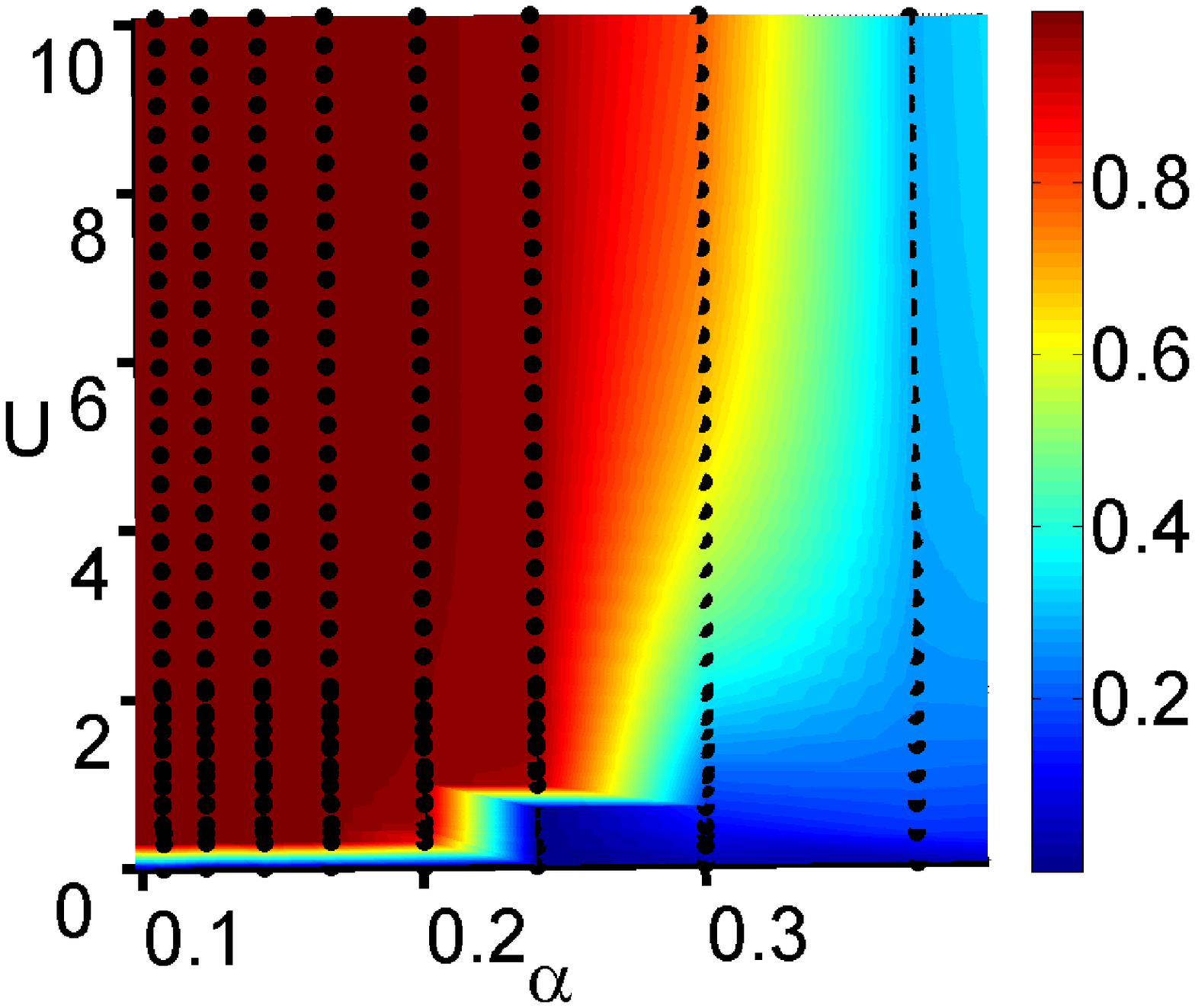}
\caption{$$}
\end{subfloat}
\hspace{0cm} \vspace{0cm}
 \begin{subfloat}
\includegraphics[width=.47\textwidth]{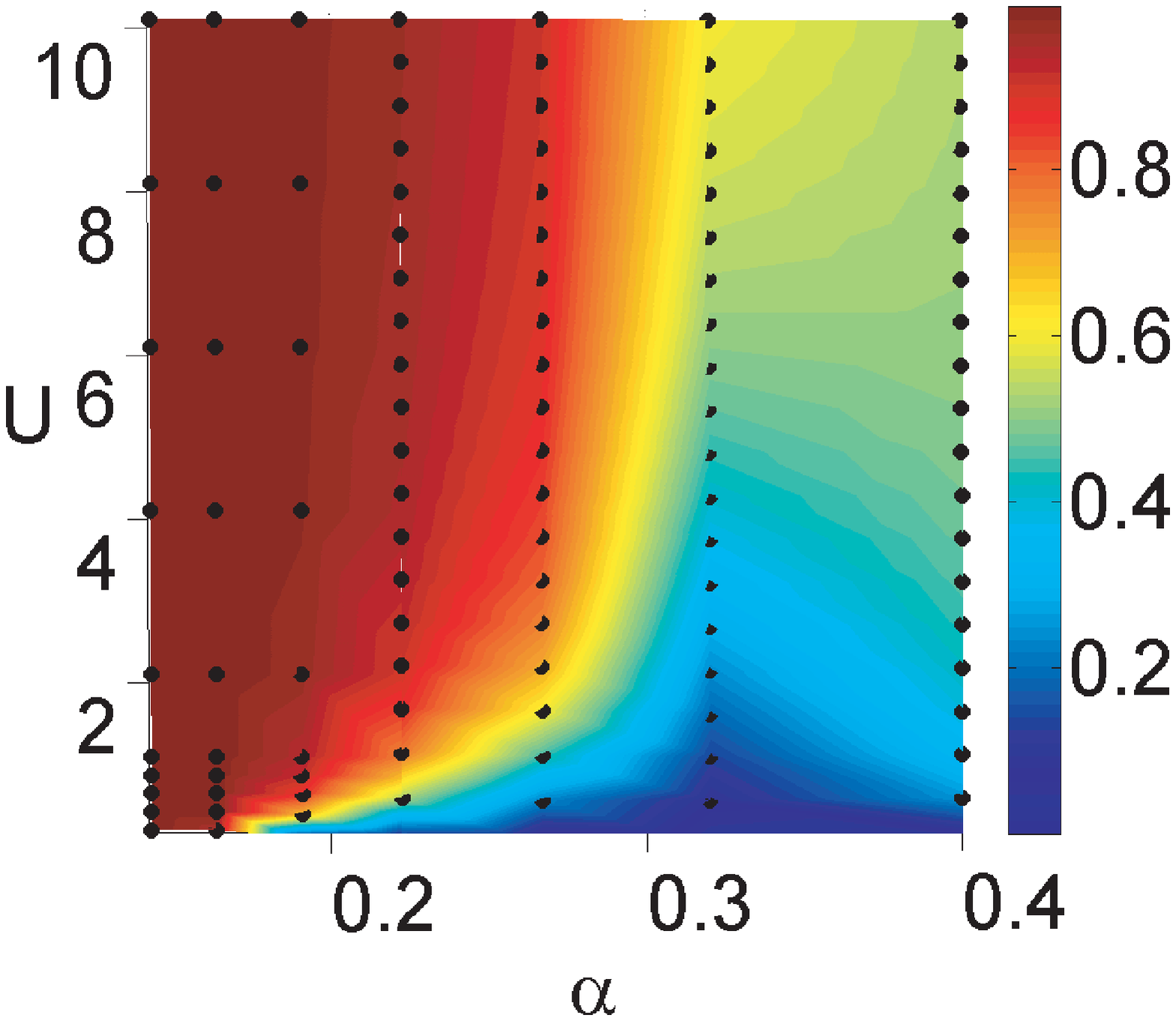}
\caption{$$}
\end{subfloat}

 \caption
{ (color online) Ground state overlap with the Laughlin wavefunction. (a) and (b)
are for 3 and 4 atoms on a lattice, respectively. $\alpha$ is varied
by changing the size of the lattice (the size in the two orthogonal
directions differ at most by unity). The Laughlin state ceases  to
be a good description of the system as the lattice nature of the
system becomes more apparent $\alpha \gtrsim 0.25$. The overlap is
only calculated at the positions shown with dots and the color
coding is an extrapolation between the points.}
\label{high-alpha-overlap}
\end{figure}

So far we have mainly considered a dilute lattice $\alpha \lesssim
0.2$, where the difference between a lattice and the continuum is
very limited. We shall now begin to investigate what happens for
larger values of $\alpha$, where the effect of the lattice plays a
significant role. Fig.~\ref{high-alpha-overlap}, shows the ground
state overlap with the Laughlin wavefunction as a function of the
strength of magnetic flux $\alpha$ and the strength of the onsite
interaction $U$. As $\alpha$ increases the Laughlin overlap is no
longer an exact description of the system since the lattice behavior
of the system is more pronounced comparing to the continuum case.
This behavior doesn't depend significantly on the number of
particles for the limited number of particles that we have
investigated $N\leq 5$. We have, however, not made any modification
to the Laughlin wave function to take into account the underlying
lattice, and from the calculations presented so far, it is unclear
whether the decreasing wave function overlap represents a change in
the nature of the ground state, or whether it is just caused by a
modification to the Laughlin wave function due the difference
between the continuum and the lattice. To investigate this, in the
next Section, we provide a  better characterization of the ground
state in terms of  Chern numbers, which shows that the same
topological order is still present in the system for higher values
of $\alpha$.

As a summary, we observe that the Laughlin wavefunction is a good
description for the case of dilute lattice ($\alpha \ll 1$),
regardless of the strength of the onsite interaction. However, the
protective gap of the ground state becomes smaller for weaker values
of interaction and in the perturbative regime $U \ll J$ is
proportional to $\alpha U$ for $\alpha\lesssim 0.2$.

\section{Chern number and topological invariance} \label{section-chern}

\subsection{Chern number as a probe  of topological order}

In the theory of quantum Hall effect, it is well understood that the
conductance quantization, is due to the existence of certain
topological invariants, so called Chern numbers. The topological
invariance was first introduced by Avron \textit{et al.}\cite{avron}
in the context of Thouless {\it et al.} (TKNdN)'s original theory
\cite{TKNdN} about quantization of the conductance. TKNdN in their
seminal work, showed that the Hall conductance calculated from the
Kubo formula can be expressed into an integral over the magnetic
Brillouin zone, which shows the quantization explicitly. The
original paper of TKNdN deals with the single-particle problem and Bloch
waves which can not be generalized to topological invariance. The
generalization to many-body systems has been done by Niu \textit{et
al.}\cite{niu85} and  also Tao \textit{et al.}\cite{tao}, by
manipulating the phases describing the closed boundary conditions on
a torus (i.e. twist angles), both for the integer and the fractional
Hall systems. These twist angles come from natural generalization of
the closed boundary condition.

To clarify the origin of these phases, we start with a single
particle picture. A single particle with charge ($q$) on a  torus of
the size $(L_x,L_y)$ in the presence of a magnetic field $B$
perpendicular to the torus surface,  is described by the Hamiltonian
\<H_s= \frac{1}{2m} \left[  \left(-i \hbar \frac{\partial}{\partial
x}-q A_x \right)^2 + \left(-i \hbar \frac{\partial}{\partial y}-q
A_y \right)^2\right],\>%
where $\vec{A}$ is the corresponding vector potential
($\frac{\partial A_y}{\partial x}-\frac{\partial A_x}{\partial
y}=B$). This Hamiltonian is invariant under the magnetic
translation,

\< t_s(\textbf{a})=e^{i \textbf{a}\cdot\textbf{k}^s/\hbar}\>%
where $\textbf{a}$ is a vector in the plane, and $\textbf{k}^s$ is the
pseudomomentum, defined by

\begin{eqnarray}
k_x^s &=& -i \hbar \frac{\partial}{\partial x}-q A_x - q By\nonumber\\
k_y^s &=& -i \hbar \frac{\partial}{\partial y}- q A_y + q Bx
\end{eqnarray}

\begin{figure}
  \centering
    \begin{subfloat}
    \label{twist}
    \includegraphics[width=.45\textwidth, height=.25\textheight,angle=0]{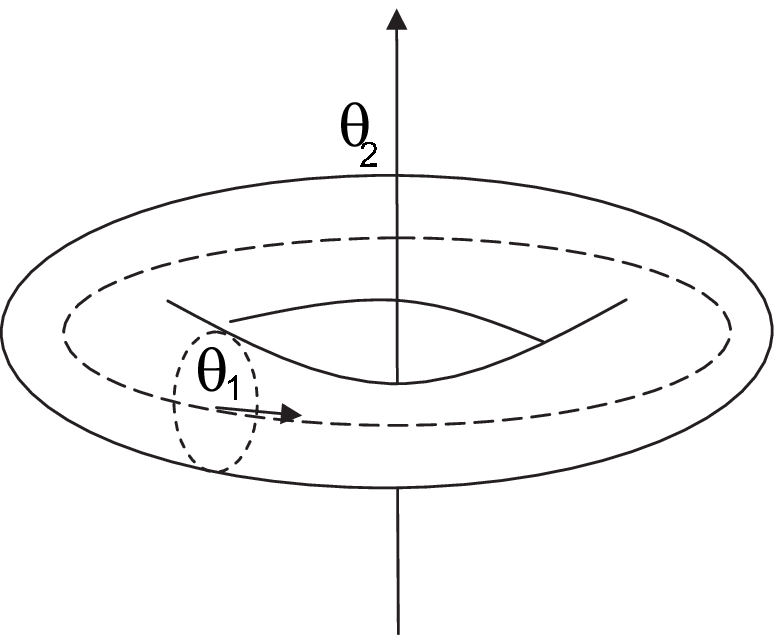}
    \caption{$$}
    \end{subfloat}
 \hspace{0cm}\vspace{0cm}
    \begin{subfloat}
    \label{gauge}
    \includegraphics[width=.45\textwidth, height=.43\textheight,angle=0]{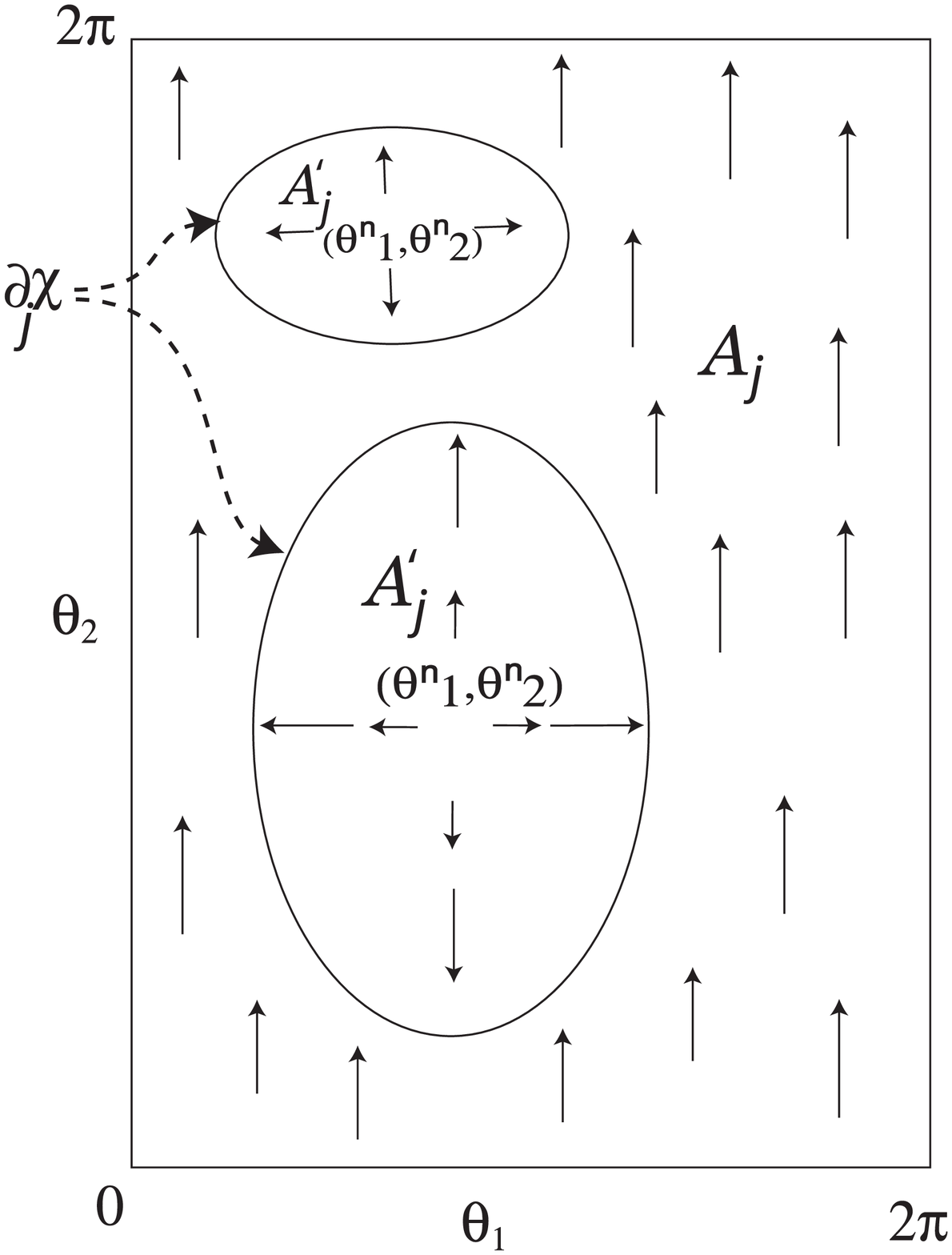}
        \caption{$$}
    \end{subfloat}

     \caption{ (a) Twist angles of the toroidal boundary condition.
In addition to the flux going through the surface there may also be
a flux inside the torus or going through the hole in the middle.
When encircling these fluxes the wave function acquire an extra
phase represented by the boundary conditions in Eq.
(\ref{twist_angles}) (b) Redefining the vector potential around the
singularities: $\mathcal{A}_j$ is not well-defined everywhere on the
torus of the boundary condition. Therefore, another vector field
$\mathcal{A'}_j$ with different definition should be introduced
around each singularity $(\theta_1^n,\theta_2^n)$ of
$\mathcal{A}_j$. $\mathcal{A}_j$ and $\mathcal{A'}_j$ are related to
each other by a gauge transformation $\chi$ and the Chern number
depends only on the loop integrals of $\chi$ around those
singularities regions, c.f., Eq.(\ref{chern_gauge}).} \label{torus}
\end{figure}

The generalized boundary condition on a torus is given by
the single-particle translation

\begin{eqnarray}
t_s(L_x\hat{x})\psi(x_s,y_s) &=&e^{i\theta_1}\psi(x_s,y_s) \nonumber \\
t_s(L_y\hat{y})\psi(x_s,y_s) &=& e^{i\theta_2}\psi(x_s, y_s) \label{twist_angles}
 \end{eqnarray}%
where $\theta_1$ and $\theta_2$ are twist angles of the boundary.
The origin of these phases can be understood by noting that the
periodic boundary conditions corresponds to the torus in Fig.
\ref{torus}(a). The magnetic flux through the surface arises from
the field perpendicular to the surface of the torus. However, in
addition to this flux, there may also be fluxes due to a magnetic
field residing inside the torus or  passing through the torus hole,
and it is these extra fluxes which give rise to the phases. The
extra free angles are all the same for all particles and all states
in the Hilbert space, and their time derivative can be related to
the voltage drops across the Hall device in two dimensions.

The eigenstates of the Hamiltonian, including the ground state will
be a function of these boundary angles $\Psi^{(\alpha)}
(\theta_1,\theta_2)$. By defining some integral form of this
eigenstate, one can introduce quantities, that do not depend on the
details of the system, but reveal general topological features of
the eigenstates.

First we discuss the simplest situation, where the ground state is
non-degenerate, and later we shall generalize this to our situation
with a degenerate ground state. The Chern number is in the context
of quantum Hall systems related to a measurable physical quantity,
the Hall conductance. The boundary averaged Hall conductance for the
(non-degenerate) $\alpha$th many-body eigenstate of the Hamiltonian
is \cite{tao, niu85}: $\sigma_{H}^\alpha=C(\alpha) e^2/h$, where the
Chern number $C(\alpha)$ is given by

\< C(\alpha)= \frac{1}{2 \pi}\int_0^{2\pi} d\theta_1 \int_0^{2\pi}
d\theta_2 (\partial_1 \mathcal{A}_2^{(\alpha)}-\partial_2
\mathcal{A}_1^{(\alpha)}) \label{eq:chern},\>
where
$\mathcal{A}_j^{(\alpha)}(\theta_1,\theta_2)$ is defined as a vector
field based on the eigenstate $\Psi^{(\alpha)} (\theta_1,\theta_2)$ on
the boundary torus $S_1 \times S_1$ by
\< \mathcal{A}_j^{(\alpha)}(\theta_1,\theta_2) \doteq i
\langle\Psi^{(\alpha)}|\frac{\partial}{\partial
\theta_j}|\Psi^{(\alpha)}\rangle. \>

It should be noted that the wave function  $\Psi^{(\alpha)}
(\theta_1,\theta_2)$ is defined up to a phase factor on the
boundary-phase space. Therefore, $\Psi^{(\alpha)}
(\theta_1,\theta_2)$ and $ e^{i f(\theta_1,\theta_2)}\Psi^{(\alpha)}
(\theta_1,\theta_2)$ are physically equivalent for any smooth
function $f(\theta_1,\theta_2)$. Under this phase change,
$\mathcal{A}_j^{(\alpha)}(\theta_1,\theta_2)$ transforms like a
gauge:

\begin{eqnarray} \mathcal{A}_j^{(\alpha)}(\theta_1,\theta_2) &\rightarrow&
\mathcal{A}_j^{(\alpha)}(\theta_1,\theta_2)  - \partial_j
f(\theta_1,\theta_2)\\ \nonumber
\Psi^{(\alpha)} (\theta_1,\theta_2) &\rightarrow& e^{i
f(\theta_1,\theta_2)}\Psi^{(\alpha)} (\theta_1,\theta_2). \label{eq:gauge_wavefunction}
\end{eqnarray}

Hence, the Chern number integral is conserved under this gauge
transformation and it encapsulates general properties of the system.
Chern numbers has been used extensively in the quantum Hall
literature, for characterization of the localized and extended
states (Ref.~\onlinecite{Sheng03} and refs. therein). In this paper,
we use the Chern number as an indicator of order in the system.
Moreover, it enables us  to characterize the ground state in
different regimes, especially where the calculation of the overlap
with the Laughlin wave function fails to give a conclusive answer.

Before explaining the method for calculating the Chern number, we
clarify some issues related to the degeneracy of the ground state.
In some systems, the ground state can be degenerate, this can be
intrinsic or it can be as a result of the topology of the system. If
the ground state is degenerate,  we should generalize the simple
integral form of Eq.\ (\ref{eq:chern}) to take into account the
redundancy inside the ground state manifold. For example, in the
case of a two-fold degenerate ground state, an extra gauge freedom
related to the relative phase between two ground states, and this
freedom should be fixed. In other words, as we change the twist
angles, we can not keep track of the evolution of both states, since
one can not distinguish them from each other. Therefore, to uniquely
determine the Chern number of the ground state(s), we should resolve
this gauge invariance, which is treated in Section
 \ref{section-impurity} and \ref{gauge-fixing}.

It important to note that the degeneracy in the non-interacting
regime is fundamentally different from the degeneracy in the
interacting case. In the non-interacting limit, the degeneracy can
be lifted by a local perturbation e.g. impurity, while in the
hardcore case, the degeneracy remains in the thermodynamic limit
\cite{wen90}. The latter degeneracy in the ground state is a
consequence of the global non-trivial properties of the manifold on
which the particles move rather than related to a symmetry breaking
which happens in conventional models e.g. Ising model. The
topological degeneracy is not a consequence of breaking of any
symmetry only in the ground state, instead it is the entire Hilbert
space which is split into disconnected pieces not related by any
symmetry transformation. With a general argument, Wen \cite{wen90b}
showed that if the degeneracy of a chiral spin system moving on a
simple torus is $k$, then it should be $k^g$ on a torus with $g$
handles (Riemann surface with genus $g$), therefore the topological
detergency is an intrinsic feature of the system. In particular, in
the context of quantum Hall effect, this multicomponent feature of
the ground state on a torus has a physical significance, while the
single component ground state on a sphere boundary condition gives
zero conductance, the torus geometry with multicomponent ground
state results in a correct conductance measured in the experiment,
since the torus boundary condition is more relevant to the
experiment. Changing twists angles of the boundary will rotate these
components into each other and gives an overall non-zero
conductance \cite{niu85}.

As studied in a recent work by M. Oshikawa and T. Senthil
\cite{senthil}, as a universal feature, it has been shown that in
presence of a gap, there is a direct connection between the
fractionalization and the topological order. More precisely, once a
system has some quasiparticles with fractional statistics, a
topological degeneracy should occur, which indicates the presence of
a topological order. Therefore, the amount of the degeneracy is
related to the statistics of the fractionalized quasiparticles e.g.
in the case of $\nu=1/2$, the two-fold degeneracy is related to 1/2
anyonic statistics of the corresponding quasiparticles. Chern number
has been also studies for spin $1/2$ system on a honeycomb lattice
\cite{kitaev2005} for identifying Abelian and non-Abelian anyons.
Bellissard {\it et al.} \cite{bellissard} studies Chern number for
disordered Fermi system using a non-commutative geometry.

To resolve the extra gauge freedom related to the two degenerate
ground states, we consider two possibilities: I. lifting the
degeneracy by adding some impurities, II. fixing the relative
phase between the two states in the ground state. Below, we explore both cases.
In the first case, we introduce some fixed impurities
to lift the degeneracy in the ground state for all values of the
twist angles. This is an artifact of the finite size of the system
which we take advantage of. In the presence of perturbation, we show
that the system has a topological order in spite of poor overlap
with the Laughlin state. In the second approach, we use a scheme recently proposed by Hatsugai \cite{varnhagen,hatsugai05} which is a generalized form for degenerate manifolds.

\subsection{Resolving the degeneracy by adding
impurities}\label{section-impurity}

In this section, we introduce some perturbation in the finite system
in form of local potentials (similar to local impurities in
electronic systems) to split the degeneracy of the ground state and
resolve the corresponding gauge invariance, which allows us to
compute the Chern number. Furthermore, the fact that we can still uniquely determine the Chern number in the presence of impurities,  shows that the system retains its topological order, even when the impurities distort its ground state wavefunction away from the Laughlin wavefunction.

In the context of the quantum Hall effect, the conventional
numerical calculation of various physical quantities such as the
energy spectrum, screening charge density profile, wave functions
overlaps, and the density-density correlation functions, can not be
used for understanding the transport properties of electrons in
the presence of  impurities (although useful for studying of isolated
impurities \cite{impurity1,impurity2}). Recently, D. N. Sheng
\textit{et al.} \cite{Sheng03} calculated the Chern number as an
unambiguous way to distinguish insulating and current carrying
states in the $\nu=1/3$ quantum Hall regime which correspond to zero
and non-zero Chern number, respectively. In this work, a weak random
disorder was used to lift the degeneracy of the ground state
(three-fold for $\nu=1/3$) for a finite number of electrons. The
energy splitting between the lowest three states then decreased with
increasing number of particles, which indicates the recovery of the
degeneracy in the thermodynamic limit. Moreover, the mobility gap
can be determined by looking at the energy at which the Chern number
drops towards zero. This energy gap is comparable to the energy
gap obtained from the experiment and it is not necessarily equal to
the spectrum gap which separates the degenerate ground state from
the excited states This shows the significance of Chern number
calculations for understanding these systems.

In a finite system, the coupling to a single-body interaction, e.g.
impurities, can lift the degeneracy and one can uniquely determine
the Chern number for the individual states by direct integration of
Eq.\ (\ref{eq:chern}). On one hand, the impurity should be strong
enough to split the energy in the ground state (in this case
$E_2,E_1$, where $E_j$ denotes the energy of the $j$th energy level)
for all values of the twist angles. On the other hand, these
impurities should be weak enough so that the energy splitting in the
ground state remains smaller than the thermodynamic gap (in this
case $E_2-E_1 \ll E_3-E_2$).

To calculate the Chern number of individual level, as mentioned in
the pervious section, we have to fix the phase of the wavefunction.
The method that we explore in this section can be considered  as a
simplified version of the general method developed by Hatsugai
\cite{hatsugai05} which we will explore in the next section.
Following Kohmoto's procedure \cite{ kohmoto}, we assume that the
ground state $\Psi (\theta_1, \theta_2)$ may be expanded for all
twist angles on a  ${\bf s}$-dimensional Hilbert discrete space
$\Psi (\theta_1, \theta_2)=(c_1, c_2,...,c_s)$. If
$\mathcal{A}_j(\theta_1,\theta_2)$ in Eq.\ (\ref{eq:chern}) is a
periodic function on the torus of the boundary condition, then by
application of Stoke's theorem, the Chern number will be always
zeros. The non-triviality (non-zero conductance in the case of
quantum Hall system) occurs because of the zeros of the wave
function, where the  phase is not well-defined. Therefore,
$\mathcal{A}(\theta_1,\theta_2)$ is not well defined everywhere and
its flux integral can be non-zero. To uniquely determine the Chern
number, we assume that the wave function and the vector field are
not defined for certain points $(\theta_1^n,\theta_2^n)$ in $S_n$
regions on the torus of the boundary condition. For simplicity, we
first discuss this procedure, in the case of a non-degenerate ground
state. For calculating the integral, we should acquire another gauge
convention for the wave function inside these $S_n$ regions, e.g. in
a discrete system, we may require an arbitrary element of the wave
function to be always real, and thereby we can define a new vector
field $\mathcal{A'}_j^{(\alpha)}(\theta_1,\theta_2)$, which is well
defined inside these regions. These two vector fields differ from
each other by a gauge transformation (Fig.~\ref{torus}):
\<\mathcal{A}_j^{(\alpha)}(\theta_1,\theta_2)-
\mathcal{A'}_j^{(\alpha)}(\theta_1,\theta_2) = \partial_j
\chi(\theta_1,\theta_2), \label{eq:1D_gauge}\>%
and the Chern number reduces to the winding number of the gauge
transformation $\chi(\theta_1,\theta_2)$ over small loops encircling
$(\theta_1^n,\theta_2^n)$, i.e. $\partial S_n$,

\<C(\alpha)= \sum_n \frac{1}{2 \pi} \oint_{\partial S_n}
\overrightarrow{\nabla} \chi \cdot d \overrightarrow{\theta}
\label{chern_gauge}.\>

The one-dimensional gauge Eq.\ (\ref{eq:1D_gauge}) should be
resolved by making two conventions. For example, in one convention
the first element and in the other the second element of the
wavefunction in the Hilbert space should be real i.e. transforming
the ground state $\Psi$ into $\Psi_{\Phi}= P \Phi=\Psi \Psi^\dagger
\Phi$ where $\Phi= (1,0,...,0)^{\dagger}$ is a ${\bf s}$-dimensional
vector and $P$ is a projection into the ground state and similarly
with the other reference vector $\Phi'= (0,1,...,0)^{\dagger}$.
Since the gauge that relates two vector fields is the same as the
one that relates the corresponding wavefunctions (similar to
Eq.(\ref{eq:gauge_wavefunction})), we can uniquely determine the
gauge transformation function $\chi$
 by evaluating $\Omega(\theta_1,\theta_2)=e^{i \chi}=\Phi^\dagger P \Phi'$.  Therefore, the Chern
number will be equal to the number of windings of $\chi$ around
regions where $\Lambda_{\phi}=\Phi^\dagger P \Phi= |c_1|^2$ is
zero. Counting the vorticities has a vigorous computational advantage over the conventional method of direct integration of Eq.\ (\ref{eq:chern}). In the direct integration, we need to choose a large number of mesh points for the boundary angles, because of the discrete approximation of derivatives in Eq.\ (\ref{eq:chern}), and this makes the calculation computationally heavy. We note that for the system on a lattice, we should  exactly diagonalize the Hamiltonian which is a sparse matrix as opposed to the continuum case where the Hamiltonian is a dense matrix residing on a smaller projected Hilbert space (lowest Landau level).

\begin{widetext}

\begin{figure}[h]
\begin{center}
\includegraphics[width=15cm,angle=0]{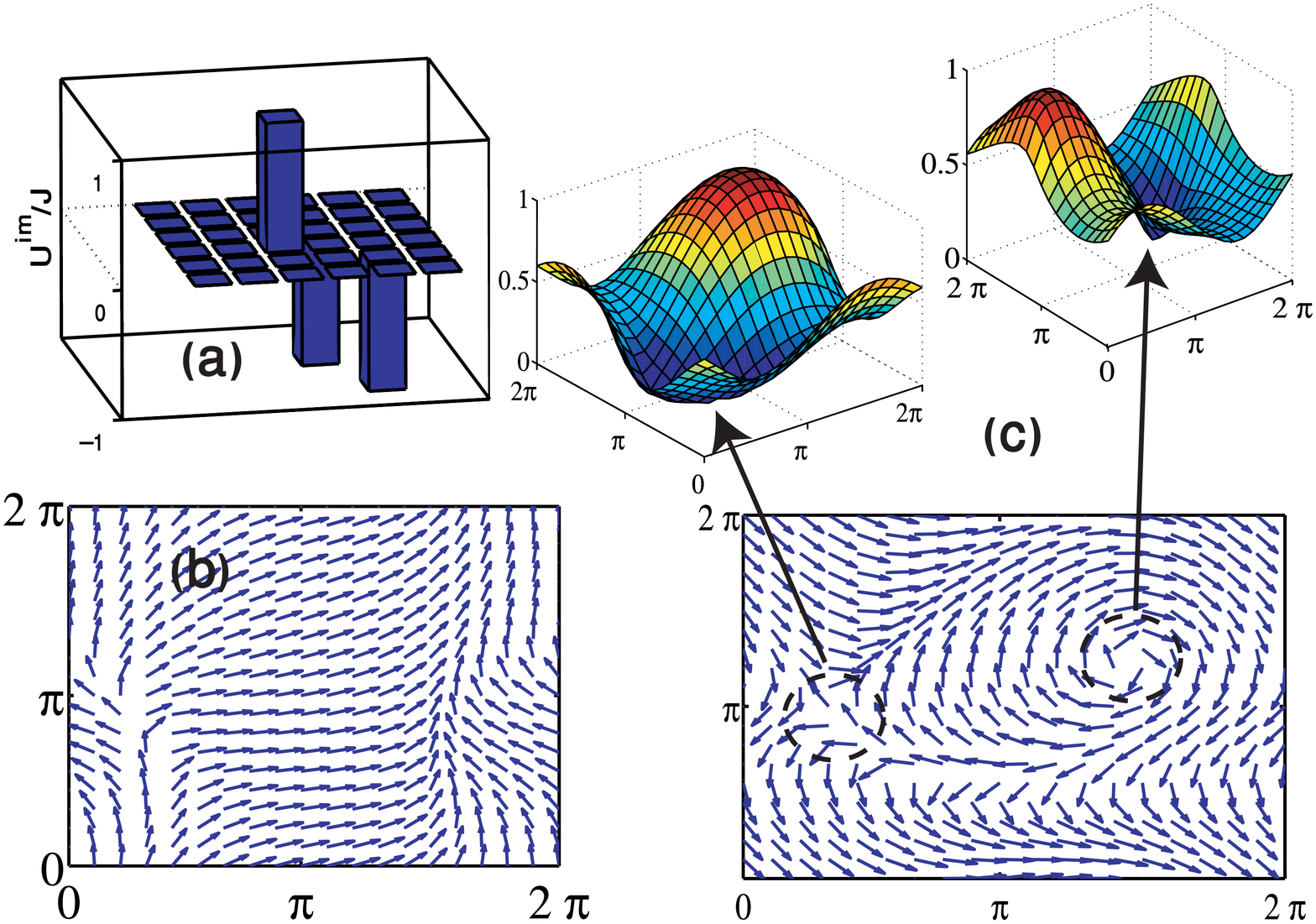}
\caption{ (color online) Chern number associated to low-lying energy
states, in the presences of impurities. Due to the impurity
potential (a), two-fold degenerate ground state splits and the
wavefunction overlap with the Laughlin state drops to $ 52 \%$ and $
65 \% $, for the first and the second energy state, respectively.
The results are for 3 atoms on a 6x6 lattice ($\alpha=0.17$) in the
hard-core limit. (b) $\Omega(\theta_1,\theta_2)$ for the first level
has no vorticity. However, for the second level, as shown in (c),
$\Omega(\theta_1,\theta_2)$ has vorticity equal to one associated
with regions where either $\Lambda_\phi$ or $\Lambda_\phi'$
vanishes. } \label{impurity}
\end{center}
\end{figure}

\end{widetext}

For removing the degeneracy, in our numerical simulations, we add a
small amount of impurity which is modeled as delta function
potentials located at random sites with a random strength, of the
order of the tunneling energy $J$. This is described by a
Hamiltonian of the form $H= \sum_i U^{im}_{i} \hat{n_i}$, where $i$
numerates the lattice site, $\hat{n}_i$ is the atom number operator
and $U^{im}_{i}$ is the strength of the impurity at site $i$.

We choose reference states $\Phi$ and $\Phi'$ to be eigenvectors of
the numerically diagonalized Hamiltonian at two different twist
angles. In Fig.\ \ref{impurity}, vorticities of $e^{i \chi}$
associated with the first and the second energy level is depicted.
It is easy to see that the Chern number associated to the two ground
states is one. Number of vortices may vary for the first and second
ground states, but their sum is always equal to one. The hard-core
limit ($U \gg J$) is very similar to the case of fractional quantum
Hall effect, which in the context of Hall systems, means a share of
$1/2$ (in $e^2/h$ unit) for each ground state \cite{tao}. When the
onsite interaction strength is small ($U<J$), the thermodynamic gap
becomes comparable to the ground state energy splitting $E_2-E_1
\sim E_3-E_2$, the Chern number can not be uniquely determined, and
the system doesn't have topological order. On the other hand, in the
limit of strong interaction ($U\gg J$), the total Chern number
associated to the ground states is equal to one, regardless of the
impurity configuration. Moreover, in the hard-core limit, although
the ground state is not described by the Laughlin wavefunction,
since it is distorted by the impurity ( in our model it can be as
low as $50\%$), the Chern number is unique and robust. This is an
indication that the topological order is not related to the
symmetries of the Hamiltonian and it is robust against arbitrary
local perturbations \cite{wen90}. These results indicate the
existence of a topological order in the system and robustness of the
ground states against local perturbations.

\subsection{Gauge fixing}\label{gauge-fixing}

The method developed in the previous section has the graphical
vortex representation for the Chern number which makes it
computationally advantageous compared to  the direct integration of
Eq.\ (\ref{eq:chern}). It can not, however, be applied directly to a
degenerate ground state, and therefore we had to introduce an
impurity potential which lifted the degeneracy. On the other hand, a
significant amount of impurity in the system may distort the energy
spectrum, so that the underlying physical properties of the lattice
and fluxes could be confounded by the artifacts due to the
impurities, especially for large $\alpha$. To address this issues,
in this section, we explore a generalized method of the previous
section based on Refs.~\onlinecite{varnhagen} and
\onlinecite{hatsugai05}, which works for a degenerate ground state.

\begin{figure}
  \centering%
  \begin{subfloat}%
\includegraphics[width=.45\textwidth,angle=0]{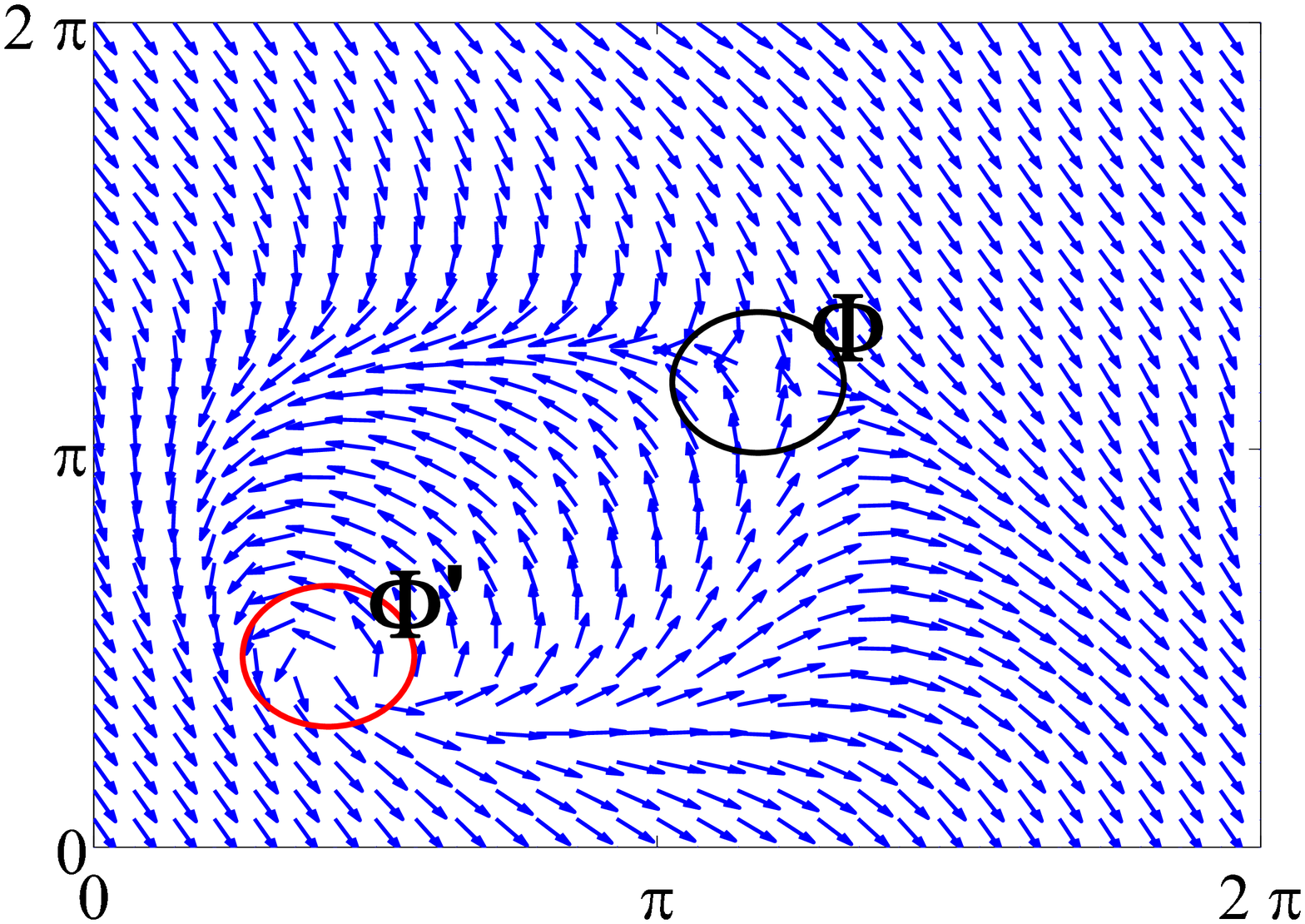}
\caption{$$}
 \end{subfloat}%
\hspace{0cm} \vspace{0cm}
  \begin{subfloat}%
\includegraphics[width=.45\textwidth,angle=0]{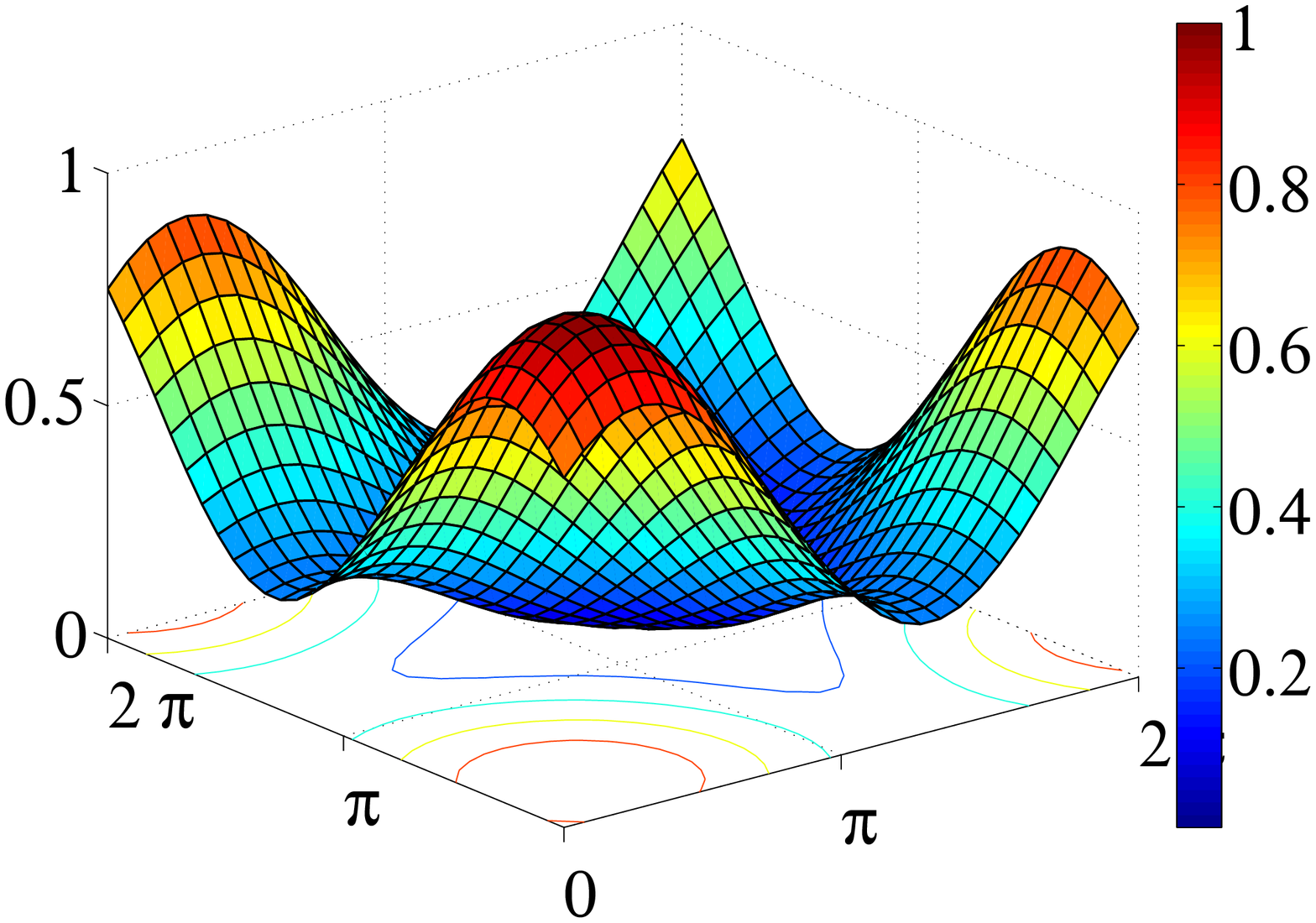}
\caption{$$}
 \end{subfloat}%
 \hspace{0cm} \vspace{0cm}
  \begin{subfloat}%
\includegraphics[width=.45\textwidth,angle=0]{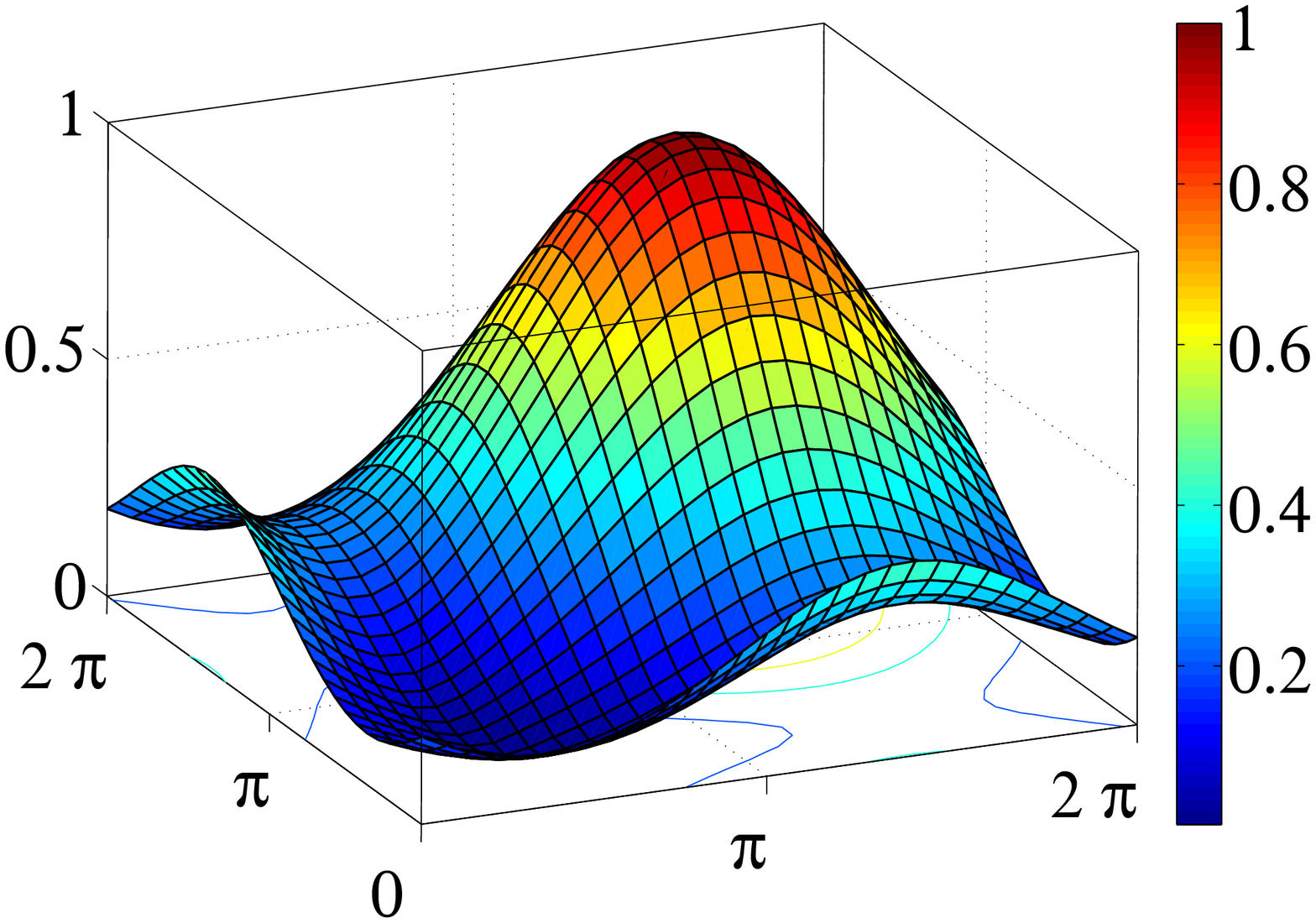}
\caption{$$}
   \end{subfloat}%

\caption{ (color online) (a) shows the argument of $\Omega(\theta_1,
\theta_2)$ as arrows for fixed $\Phi$ and $\Phi'$. (b) and (c):
surface plots of ${\rm det}\Lambda_\Phi$ and ${\rm det}
\Lambda_\Phi'$ (blue is lower than red). $\theta_1$ and $\theta_2$
changes from zero to $2\pi$. These plots have been produced for 3
atoms with $N_\phi=6$  ($\alpha=0.24$) in the hard-core limit on a
5x5 lattice. The total vorticity corresponding to each of the
reference wave functions ($\Phi$ or $\Phi'$) indicates a Chern
number equal to one.}
  \label{omega}
\end{figure}

By generalizing the Chern number formalism for a degenerate ground state manifold, instead of having a single vector
field $ \mathcal{A}_j^{(\alpha)}(\theta_1,\theta_2) $, a tensor
field $ \mathcal{A}_j^{(\alpha, \beta)}(\theta_1,\theta_2) $ should
be defined, where $\alpha, \beta= 1,2, ..., q$ for a ${\bf q}$-fold degenerate
ground state

\< \mathcal{A}_j^{(\alpha,\beta)}(\theta_1,\theta_2) \doteq i
\langle\Psi^{(\alpha)}|\frac{\partial}{\partial
\theta_j}|\Psi^{(\beta)}\rangle \>

Similar to the non-degenerate case,
when $\mathcal{A}_j^{(\alpha,\beta)}$ is not defined, a new gauge convention should be
acquired for the regions with singularities. This gives rise to a
tensor gauge transformation $ \chi^
{(\alpha,\beta)} (\theta_1,\theta_2)$ on the border of these regions

\<\mathcal{A}_j^{(\alpha,\beta)}(\theta_1,\theta_2)-
\mathcal{A'}_j^{(\alpha,\beta)}(\theta_1,\theta_2) = \partial_j
\chi^{(\alpha,\beta)}(\theta_1,\theta_2).\>

Following Hatsugai's proposal\cite{hatsugai05} for fixing the ground
state manifold gauge, we take two reference multiplets $\Phi$ and
$\Phi'$ which are two arbitrary  ${\bf s} \times {\bf q}$ matrices;
$q$ is the ground state degeneracy (equal to 2 in our case). In our
numerical simulation, we choose the multiplets to be two sets of
ground state at two different twist angles far from each other, e.g.
$(0,0)$ and $(\pi,\pi)$. We define an overlap matrix as
$\Lambda_\phi=\Phi ^\dagger P \Phi$ where $P= \Psi \Psi^\dagger$ is
again the projection into the ground state multiplet, and consider
the regions where ${\rm det} \Lambda_\Phi$ or ${\rm det}
\Lambda_{\Phi'}$ vanishes (similar to zeros of the wave function in
the non-degenerate case). Hence, the Chern number for ${\bf q}$
degenerate states, will be equal to the total winding number of
$\textrm{Tr}~ \chi^ {(\alpha,\beta)}$ for small neighborhoods,
$S_n$, in which ${\rm det} \Lambda_\Phi$ vanishes

\begin{figure}
  \centering%
  \begin{subfloat}%
\includegraphics[width=.45\textwidth,angle=0]{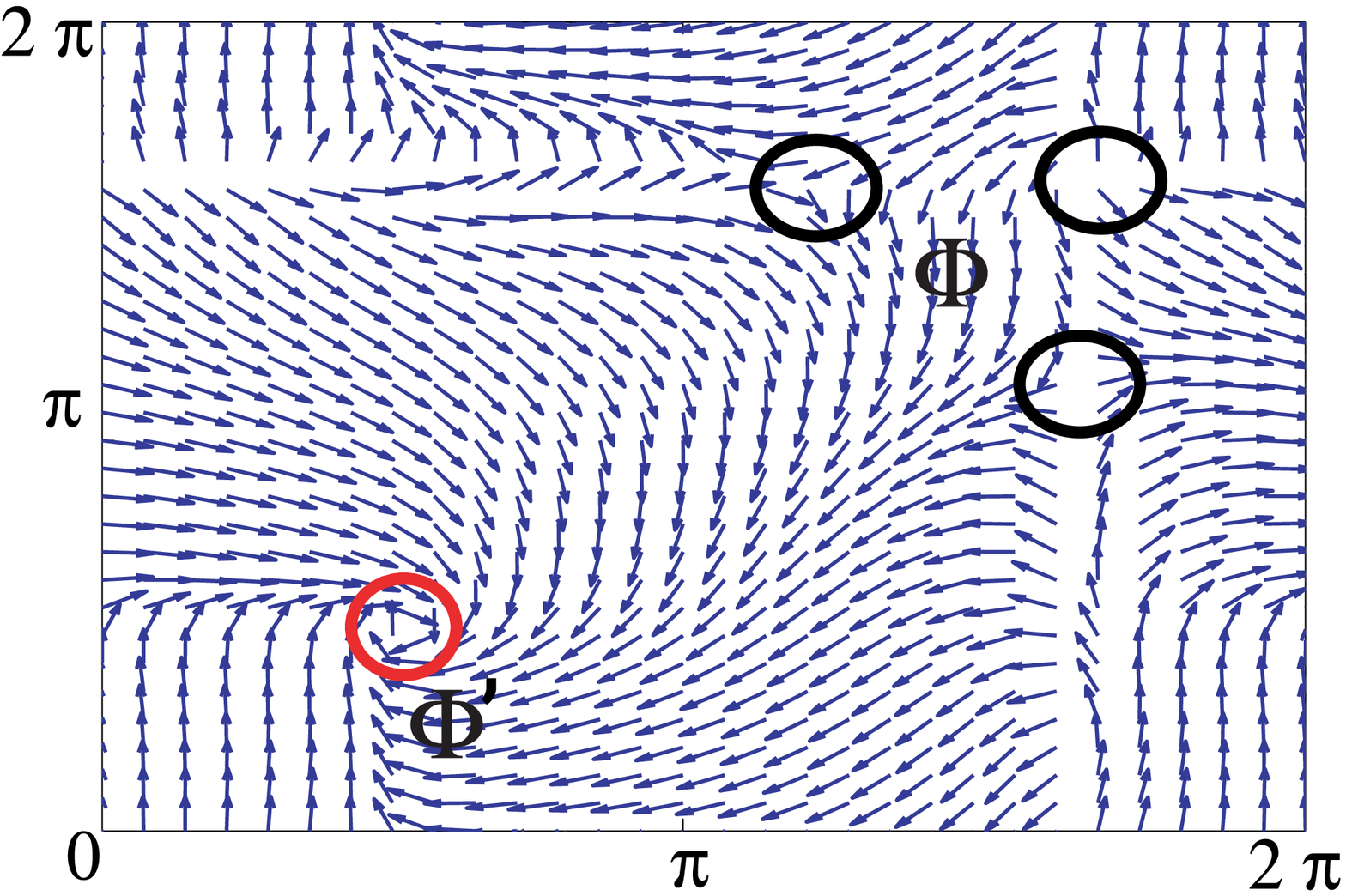}
 \end{subfloat}%

  \caption{ (color online) $\Omega(\theta_1, \theta_2)$ for fixed $\Phi$
and $\Phi'$. $\theta_1$ and $\theta_2$ changes form zero to $2\pi$.
This plot has been produced for 4 atoms with $N\phi=8$ in the
hard-core limit on a 5x5 lattice ($\alpha=0.32$). Although, there
are more vortices here compared to Fig.~\ref{omega}, the total
vorticity corresponding to each of the trial functions ($\Phi$ or
$\Phi'$) indicates a Chern number equal to one.}
  \label{omega-4}
\end{figure}

\begin{eqnarray}
C(1,2,...,q)&=&\sum_n \frac{1}{2 \pi} \oint_{\partial S_n}
\overrightarrow{\nabla} \textrm{Tr}~\chi^ {(\alpha,\beta)}
 \cdot d \overrightarrow{\theta}
\end{eqnarray} %
which is the same as the number of  vortices of  $\Omega(\Phi,
\Phi')={\rm det}(\Phi^\dagger P \Phi')$. It should be noted that the
zeros of det $\Lambda_\Phi$ and det$\Lambda_\Phi'$ should not
coincide in order to uniquely determine the total vorticity. In
Fig.~\ref{omega}, we have plotted $\Omega$, ${\rm det}\Lambda_\Phi$,
and ${\rm det}\Lambda_{\Phi'}$, found by numerical diagonalization
of the Hamiltonian for a mesh ($30 \times 30 $) of winding angles
$\theta_1$ and $\theta_2$. In this figure, the Chern number can be
determined be counting the number of vortices and it is readily seen
that the winding number is equal to one for the corresponding zeros
of ${\rm det} \Lambda_\Phi$ (or ${\rm det} \Lambda_{\Phi'}$).

\begin{table}
\begin{center}
\begin{tabular}{|c|c|c|c|c|c|c|c|}   \hline
{\em Atoms  }      & {\em Lattice } &      {\em $\alpha$}   & {\em Chern/state}& Overlap  \\
\hline  3           &    6x6  &    .17      &    1/2 &  0.99 \\
\hline  4           &    6x6  &    .22       &    1/2 &  0.98 \\
\hline  3           &    5x5  &    .24       &1/2 & 0.98 \\
\hline  3           &    4x5  &    .3         &1/2 & 0.91 \\
\hline  4           &    5x5 &    .32        &    1/2 & 0.78\\
\hline  3           &    4x4 &    .375    & 1/2 & 0.29\\
\hline
\end{tabular}
\end{center}
\caption{Chern Number for different configurations in the hard-core
limit for fixed filling factor $\nu=1/2$. The Laughlin state overlap
is shown in the last column. although it deviates from the Laughlin
state. Although the ground state deviates from the Laughlin state,
the Chern number remains equal to one half per state before reaching
some critical $\alpha_c \simeq 0.4$ where the energy gap vanishes. }
\label{chern}
\end{table}

We have calculated the Chern number for fixed $\nu=1/2$ and
different $\alpha$'s by the approach described above. The result is
shown in Table \ref{chern}. For low $\alpha \ll 1$, we know from
Sec. \ref{section-no_hardcore} that the ground state is the
Laughlin state and we expect to get a Chern number equal to one. For
higher $\alpha$, the lattice structure becomes more apparent and the
overlap with the Laughlin state decreases. However, in our
calculation, the ground state remains two-fold degenerate and it
turns out that the ground state Chern number tends to remain equal
to one before reaching some critical $\alpha_c \simeq 0.4$.
Hence, also in this regime
we expect to have similar topological order and fractional
statistics of the  excitations above these states on the lattice.

\begin{figure}
  \centering%
  \begin{subfloat}%
\includegraphics[scale=.4,angle=0]{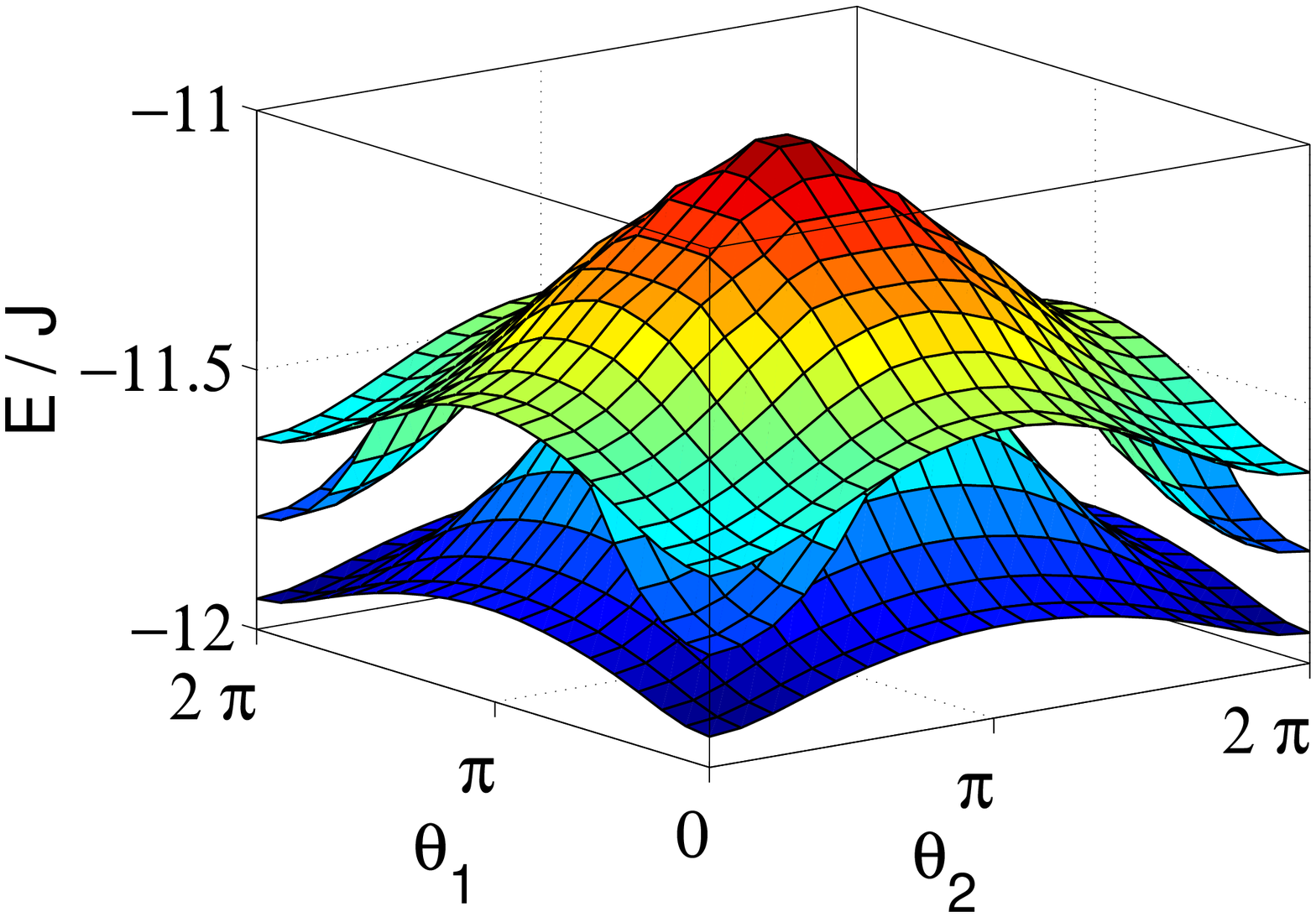}
\caption{$$}
 \end{subfloat}%
  \hspace{0cm}%
  \begin{subfloat}%
\includegraphics[scale=.4,angle=0]{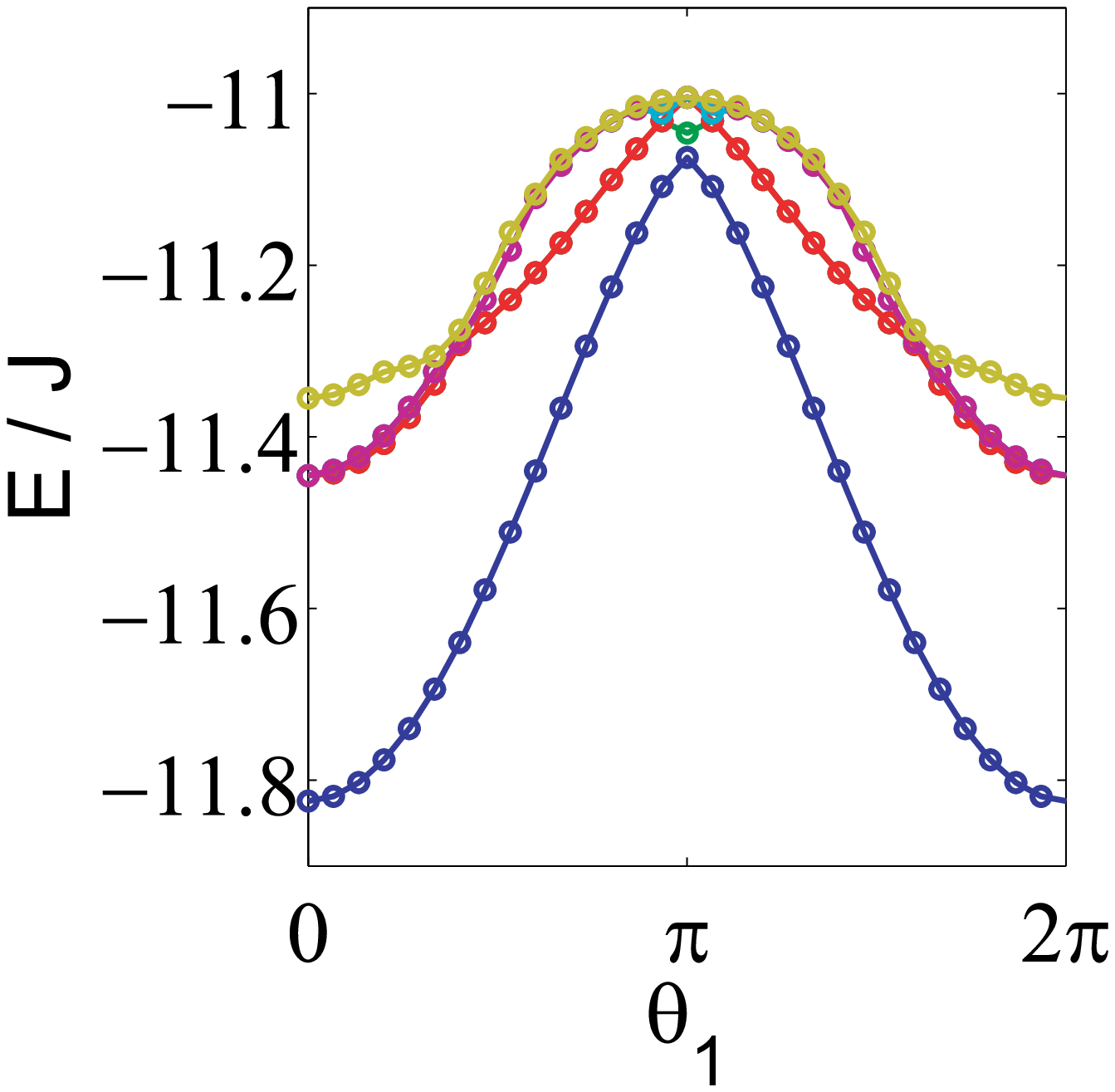}
\caption{$$}
  \hspace{0cm}%
 \end{subfloat}%
\caption{ (color online) Low-lying energy levels as a function of twist angles. For
high $\alpha$ the degeneracy of the ground state is a function of
twist angles. The shown results are for 5 atoms on a 5x5 lattice
i.e. $\alpha=0.4 $ (a) shows first three energy manifolds as a
function of the toroidal boundary condition angles, (b) shows a
cross section of (a) at $\theta_2=\pi$ for seven lowest  energy
levels. The first and the second energy levels get close to each
other at $\theta_1=\theta_2=\pi$. }
  \label{level-crossing}
\end{figure}

For the arguments above to be applicable, it is essential that we can uniquely identify a two-fold degenerate ground state which is well separated from all higher lying states.
For higher flux densities, $\alpha > \alpha_c$, the two-fold
ground state degeneracy is no longer valid everywhere on the torus
of the boundary condition. In this regime, the issue of degeneracy
is more subtle, and the finite size effect becomes significant. The
translational symmetry argument \cite{haldane85}, which was used in
the Section \ref{section-no_hardcore}, is not  applicable on a lattice and as pointed out by Kol
\textit {et al}. \cite{Kol}  the degeneracy of the ground state may vary periodically with the system size. Some of the gaps which appear in the calculation may be due to the finite size and vanish in the thermodynamic limit, whereas others may represent real energy gaps which are still present in the thermodynamic limit.
To investigate this, we study the ground state degeneracy as a function of boundary
angles ($\theta_1, \theta_2$) which are not physical observable and
therefore the degeneracy in thermodynamic limit should not depend on
 their value. In particular, Fig.~\ref{level-crossing} shows the
energy levels of five particles at $\alpha=0.4$ for different values
of the twist angles. The first and the second level are split at
($\theta_1=\theta_2=0$), while they touch each other at
($\theta_1=\theta_2=\pi$). We have observed similar behavior  for
different number of particles and lattice sizes e.g. 3 and 4 atoms at
$\alpha=0.5$. In this case, the system seems to not have a two-fold degeneracy. Therefore, the ground state enters a different regime which is a subject for further
investigation.

For having the topological order, it is not necessary to be in the
hard-core limit. Even at finite interaction strength $U \sim J
\alpha $, we have observed the same topological order with the help
of the Chern number calculation. If $U$ gets further smaller, the
energy gap above the ground state diminishes (as seen in Sec.~\ref{section-no_hardcore}) and the topological order disappears.

We conclude that the Chern number can be unambiguously calculated
for the ground state of the system in a regime where Laughlin's description is
not appropriate for the lattice. The non-zero Chern number of a
two-fold degenerate ground state, in this case equal to one half per state, is a direct indication of the
topological order of the system.

\section {Extension of the Model}

In Sections \ref{section-no_hardcore} and \ref{section-chern} above, we have investigated the conditions under which the fractional quantum Hall effect may be realized for particles on a lattice. The motivation for this study is the possibility to generate the quantum Hall effect with ultra cold atoms in an optical lattice but the results of these sections are applicable regardless of the method used to attain this situation. In this and the following sections, we investigate some questions which are of particular relevance to ultra cold atoms in an optical lattice. First, we introduce a long range, e.g., dipole-dipole, interaction  which turn out to increase the energy gap and thereby stabilizes the quantum Hall states. We then turn to the case of $\nu=1/4$ and show that in order to realize this state,  it is essential to have some kind of long range interaction.

\subsection{Effect of the long-range interaction} \label{section-dipole}
In an experimental realization of the quantum Hall effect on a
lattice, it is desirable to have as large an energy  gap as possible in order
to  be insensitive to external perturbations. So far, we have
studied effect of the short range interaction, and we have shown
that the gap increases with increasing interaction strength, but the
value of the gap saturates when the short range interaction becomes
comparable to the tunneling energy $J$.

In this section, we explore the possibility of increasing the gap by
adding a long-range repulsive dipole-dipole interaction to  the
system. Previously, such dipole-dipole interaction has also been studied in Ref.\\onlinecite{cooper2005}  as a method  to achieve Read-Rezayi states \cite{read_rezayi99}  of rapidly rotating cold trapped atoms for $\nu=3/2$ and as a means to realize fractional Quantum Hall physics with Fermi gases \cite{baranov}. The dipole-dipole (magnetic or electric) interaction is
described by the Hamiltonian: \<H_{\rm d-d} = U_{dipole} \sum_{1\leq
i<j\leq N} \frac{{\bf p}_i\cdot{\bf p}_j - 3({\bf n}_{ij}\cdot{\bf
p}_i)({\bf n}_{ij}\cdot{\bf p}_j)}{|{\bf r}_i-{\bf r_j}|^3}\> where
${\bf n}_{ij}=({\bf r}_i-{\bf r}_j)/|{\bf r}_i-{\bf r}_j|$.
${\bf p}_i$ are unit vectors representing the permanent dipole
moments and the position vectors ${\bf r}_i$ are in units of the lattice spacing $\bf a$. For
simplicity, we assume that all dipoles are polarized in the
direction perpendicular to the plane. With time independent dipoles, the strength of the interaction is given by $ U_{dipole} = \frac{\mu_0 \mu^2}{4 \pi a^3}$ (or $\frac{\wp^2}{4 \pi \epsilon_0 a^3} $) where ${\bf {\mu}}$'s ($\bf \wp$'s) are the permanent magnetic (electric) dipole moment. Static dipoles will thus give repulsive interaction  $ U_{dipole} >0$, but experimentally time varying fields may be introduced which effectively change the sign of the interaction\cite{giovanazzi}. For completeness,  we shall therefore both investigate positive and negative  $ U_{dipole}$, but the repulsive interaction corresponding to static dipoles will be the most desirable situation since it stabilizes the quantum Hall states.

Experimentally the dipole-dipole interaction will naturally be present in the recently realized  Bose-Einstein condensation of Chromium\cite{griesmaier} which has a
large magnetic moment. However, for a lattice realization, polar
molecules which have strong permanent electric dipole moments is a more  promising candidate. For typical polar molecules with the electric moment $\wp \sim 1$ Debye, on a lattice with spacing $\bf{a}$$~\sim 0.5 \mu$m, $U_{dipole}$ can be up to few kHz, an order of magnitude greater than the typical tunneling  $  J/2 \pi\hbar$ which can be few hundreds of Hz \cite{greiner}.
\begin{figure}
  \centering%
  \begin{subfloat}%
\includegraphics[scale=.30,angle=0]{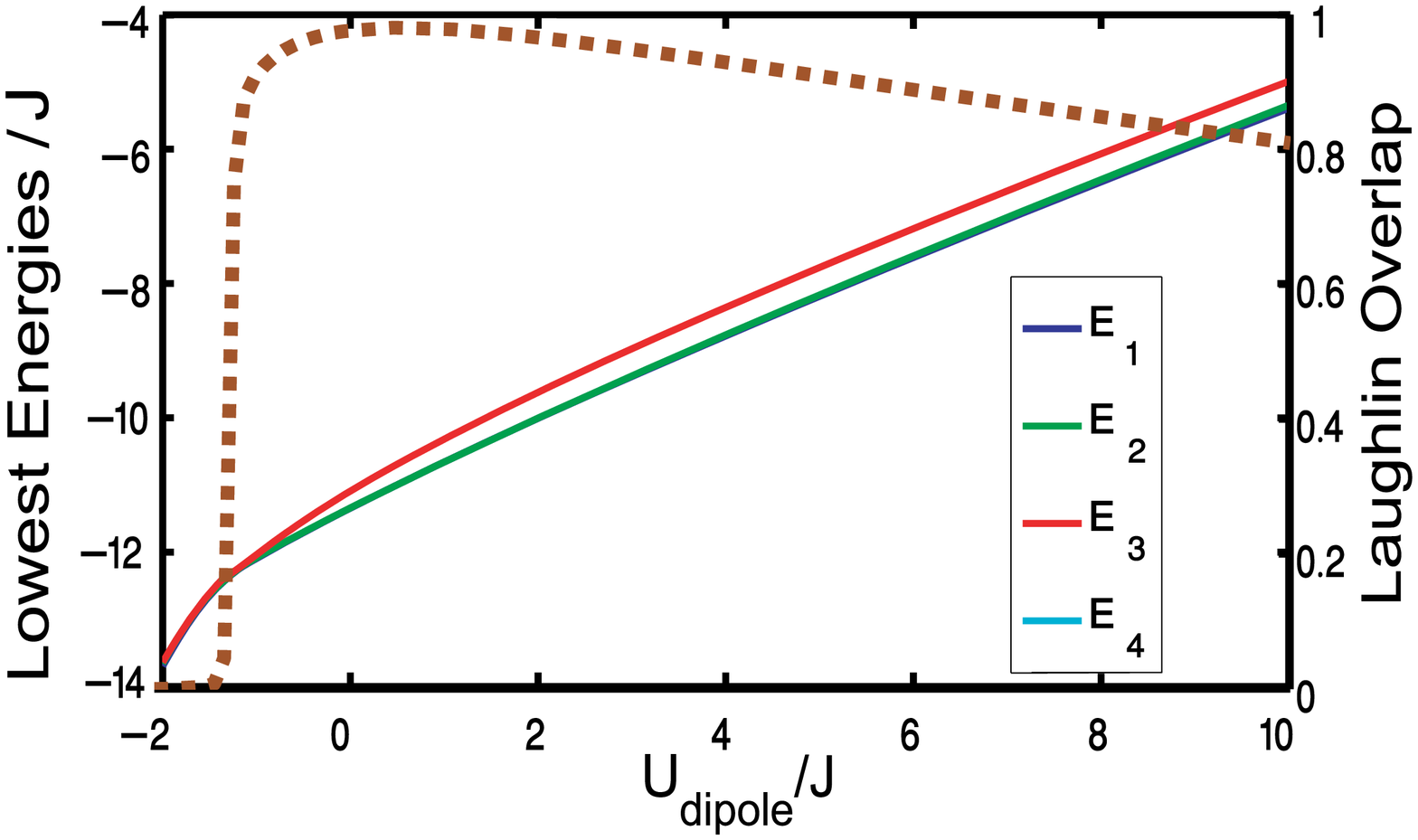}
\caption{$$}
 \end{subfloat}%
  \hspace{0cm}%
  \begin{subfloat}%
\includegraphics[width=.40\textwidth,angle=0]{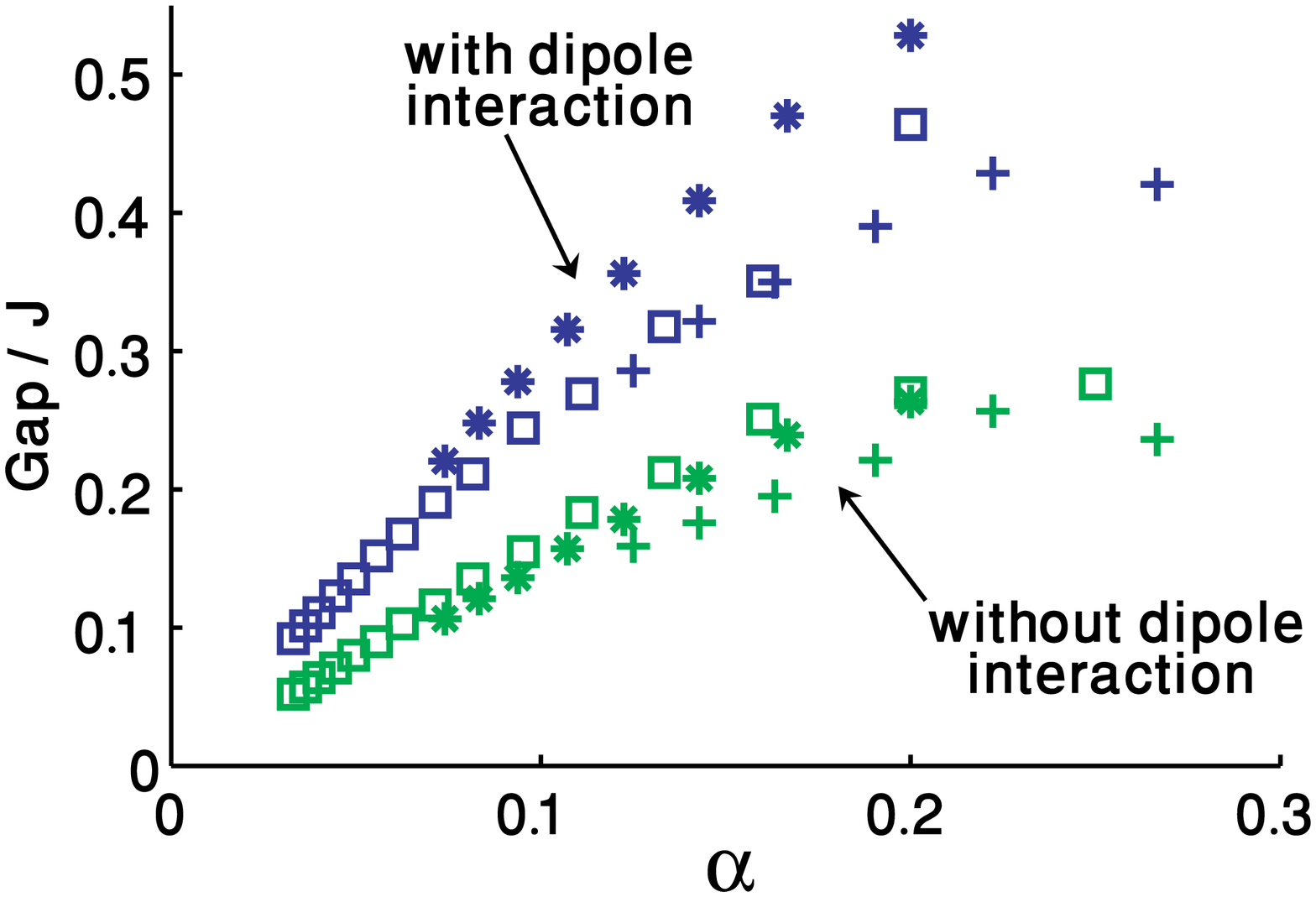}
\caption{$$}
  \hspace{0cm}%
 \end{subfloat}%

\caption { (color online) (a) The overlap of the ground state with the Laughlin
wavefunction (dashed lines) and four low lying energies of the
system (solid lines) versus the dipole-dipole interaction for four
atoms on a 6x6 lattice. (b) Gap enhancement for a fixed repulsive
dipole-dipole interaction strength $U_{dipole}=5 J$ versus $\alpha$.
The results are shown for $N = 2 (\square)$, $N = 3 (\ast)$, $N = 4
(+)$  } \label{dipole}
\end{figure}

To study the effect of the dipole-dipole interaction, we again
numerically diagonalize the Hamiltonian for a few hardcore bosons
($U\gg J$), in the dilute regime  $\alpha \leqslant
0.3$, while varying strength of the dipole-dipole interaction.  The
results of this simulation is shown in  Fig.~ \ref{dipole} (a). After
adding the dipole interaction, the ground state in a dilute lattice
remains two-fold degenerate, since the interaction only depends on
the relative distance of the particles and keeps the center of mass
motion intact.  If the dipole interaction becomes too strong $({\bf
U}_{dipole} \gg J)$, the ground state wave function deviates from
the Laughlin wavefunction, but the topological order remains the
same as for the system without dipole interaction. We verified this by
calculating the Chern number as explained in Sec.~\ref{section-chern} for different values of the dipole-dipole
interaction strength ${\bf U}_{dipole}$ and observed that the total
Chern number of the two-fold degenerate ground state is equal to
one. Moreover, as it is shown in Fig.~ \ref{dipole} (b) adding such
an interaction can increase the gap: the lower curve corresponds to
the hard-core limit discussed in the previous work \cite{sorensen}
and the upper curve corresponds to the system including the
dipole-dipole interaction. This enhancement varies linearly with the flux
density $\alpha$ in a dilute lattice and doesn't depend on the
number of particles and consequently, it is expected to behave
similarly  in the thermodynamic limit.

One of the impediment of the experimental realization of Quantum
Hall state is the smallness of the gap which can be improved by
adding dipole-dipole interaction. In this section, we showed that
this improvement is possible and moreover, by Chern number
evaluation, we verified that adding dipole interaction doesn't
change the topological behavior of the ground state manifold.

\subsection{The case of $\nu=1/4$} \label{nu1/4}

So far we have concentrated on the case of $\nu=1/2$. In this section, we
briefly investigate the case of $\nu=1/4$. It is expected that the
Laughlin wavefunction remains a good description for the ground
state of a bosonic system for any even $q$, where $\nu=1/q$.
Following Haldane's argument\cite{ haldane85}, due to the center of
mass motion, the degeneracy of the ground state is expected to be
$q$-fold on a torus. Similar to the case of $\nu=1/2$, the Laughlin
wavefunction should be a suitable description for any $q$ provided
that the magnetic field is weak so that we are close to the
continuum limit, i.e. $\alpha\ll 1$. Also the Chern number is
expected to be equal to one for the q-fold degenerate ground state,
which in the context of quantum Hall effect means a share of $1/q$
of the conduction quantum $e^2/h$ for each state in the q-fold
degenerate ground state manifold.

\begin{figure}[t]
\includegraphics[width=.40\textwidth,angle=0]{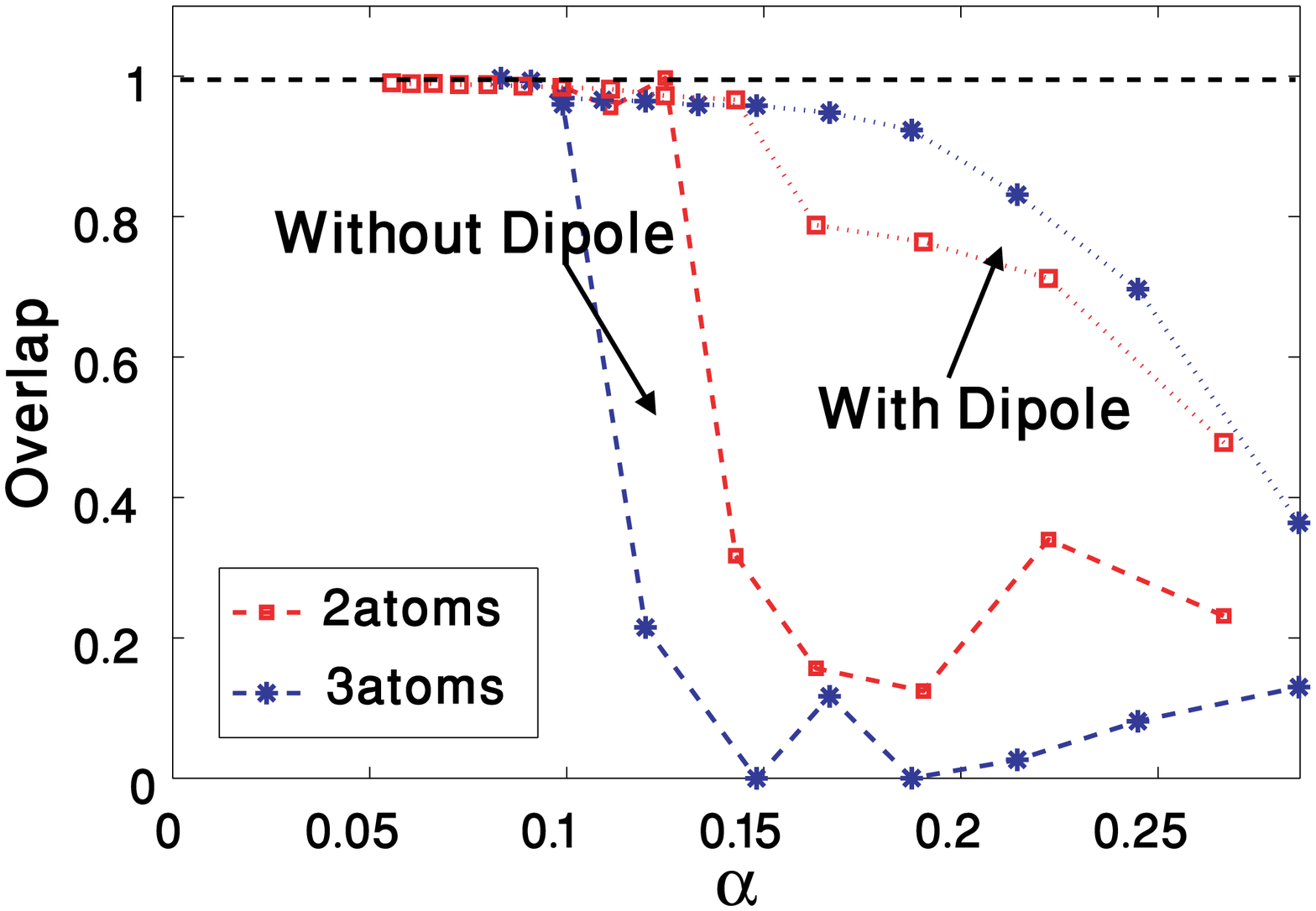}

 \caption
{ (color online) The overlap of the first four low-lying energy states with the Laughlin wavefunction
for the  case of $\nu=1/4$ on a torus. The dashed (dotted) line shows the
overlap for the system without (with) dipole interaction
($U_{dipole}=5J$). The Laughlin state is only a good description for a
more dilute lattice $\alpha \lesssim 0.1 $ compared to $\nu=1/2$.
The dipole interaction stabilizes the system and the overlap is more significant for higher values of $\alpha \lesssim 0.2$.} \label{overlap_laughlin_1_4}
\end{figure}

We have done both overlap and the Chern number calculation to check
those premises. In the case of $\nu=1/4$, significant overlap occurs
at low $\alpha \lesssim 0.1$. The average wave function overlap of
four lowest energy eigenstates with the Laughlin wavefunction is
depicted in figure \ref{overlap_laughlin_1_4}, where we have used a
generalization of the Laughlin wavefunction for periodic boundary
conditions similar to Eq. (\ref{laughlin-torus}) \cite{read96}.

We observe that the Laughlin wavefunction is a reliable description
of the system with $\nu=1/4$ but only for much more dilute lattices
($\alpha \lesssim 0.1$) compared to $\nu=1/2$ where significant
overlap occurs for $\alpha \lesssim 0.3$. Contrary to $\nu=1/2$,
where the gap is a fraction of the tunneling energy $J$, the gap for
$\nu=1/4$ between the 4-fold degenerate ground state and the next
energy level is infinitesimal. The reason for the vanishing gap can
be understood in the continuum limit from the argument put forward
in Ref.~\onlinecite{Jaksch06}: as noted previously the Laughlin
wavefunction is an exact eigenstate of the Hamiltonian with an
energy per particle equal to the lowest Landau level energy. The
energy of the $m=4$ state is thus equal to the $m=2$ state. It thus
costs a negligible energy to compress the $\nu=1/4$ state to the
$\nu=1/2$ state, and therefore there is no gap. In an external trap
the system will always choose the highest density state which is the
$\nu=1/2$ state. Note, however that this argument only applies to
short range interactions. For long range interactions, we expect to
see a non vanishing gap.

Even though that with short range interactions, the gap is very small in our numerical calculations, it is still sufficiently large that it allows us to unambiguously
determine the Chern number for the ground state manifold as
described in Sec.~\ref{gauge-fixing}. As expected the calculation
shows that the Chern number is equal to one corresponding to
a four-fold degenerate ground state consistent with the generalization
of the fermionic case in the fractional quantum Hall theory
\cite{niu85,tao}. In Fig.~ \ref{overlap_laughlin_1_4}, the overlap
of the first four lowest energy state with the Laughlin
wavefunction is depicted.  In the absence of the dipole interaction,
the ground state overlap is significant only for $\alpha
\lesssim 0.1$, however, by adding a moderate dipole interaction
($U_{dipole}=5J$), the overlap becomes more significant for a larger
interval of the flux density, i.e. $\alpha \lesssim 0.25$.
This is due to the fact that state with lower density become more
favorable in the presence of a long-range repulsive interaction.

We observed that adding a dipole interaction would lead to an improvement of the gap for $\nu=1/2$ and make the Laughlin overlap more significant for larger interval of magnetic field strength $\alpha$ in the case of $\nu=1/4$. Therefore this long-range interaction can be used as an  tool for stabilizing the ground state and make the realization of these quantum states,  experimentally more feasible.

\section{Detection of the Quantum Hall state} \label{bragg}

In an experimental realization of the quantum Hall states, it is
essential to have an experimental probe which can verify that the
desired states were produced. In most experiments with cold trapped
atoms, the state of the system is probed by releasing the atoms from
the trap and imaging the momentum distribution. In Ref.~\onlinecite{sorensen}, it was shown that this technique provide some
indication of the dynamics in the system. This measurement
technique, however, only provides limited information, since it only
measures  the  single particle density matrix, and provides no
information about the correlations between the particles.
In Refs.~\onlinecite{Jaksch06} and \onlinecite{Ehud} a more advanced measurement techniques were
proposed, where the particle correlation is obtained by looking at
the correlations in the expansion images. In this section, we study
Bragg scattering as an alternative measurement strategy which
reveals the excitation spectrum of the quantum system.  In Ref.~\onlinecite{bhat07} bosonic quantum Hall system responses to a perturbative potential is studied. We focus on Bragg
scattering where two momentum states of the same internal ground
state are connected by a stimulated two-photon process
\cite{ketterle}. By setting up two laser beams with frequencies
$\omega_1$ and $\omega_2$ and wave vectors $\vec{k_1}$ and
$\vec{k_2}$ in the plane of the original lattice, a running optical
superlattice will be produced with   frequency $\omega_1-\omega_2$
and  wave vector $\vec{k_1}-\vec{k_2}$. (Both frequencies $\omega_1$ and $\omega_2$ should be
close to an internal electronic dipole transition in the atoms). The
beams should be weak and sufficiently
detuned so that direct photon transitions are negligible,
i.e. $\mathcal{E}_1,~ \mathcal{E}_2,~ \gamma \ll \omega_1-\omega_{0},
~\omega_2-\omega_{0}$, where $\omega_{0}$ is the frequency of the
transition, $\gamma$ is the corresponding spontaneous decay rate and
$\mathcal{E}_1, \mathcal{E}_2$ are the Rabi frequencies related to the
laser-atom coupling. In this perturbative regime, the inelastic
scattering of photons will be suppressed; the atom starts from the
internal ground state, absorbs one photon from e.g. beam 1 by going
to a virtual excited state and then emits another photon into beam 2
by returning to its internal ground state. After this process, the
system has acquired an energy equal to $\hbar\omega=\hbar
(\omega_1-\omega_2)$ and a momentum kick equal to
$\vec{q}=\vec{k_1}-\vec{k_2}$. Therefore, the overall effect of the
recoil process is a moving AC Stark shift as a perturbing moving
potential, and the effective Hamiltonian represents the exchange of
the two-photon recoil momentum and the energy difference to the
system and is proportional to the local density i.e.
$H\propto \rho(r)
e^{-i(\omega_1-\omega_2)t+i(\vec{k_1}-\vec{k_2}).\vec{r}}+c.c.$.

This process can be used to probe density fluctuations of the system
and thus to measure directly the dynamic structure factor
$S(q,\omega)$ and static structure factor $S(q)$. This kind of
spectroscopy has been studied for a BEC in a trap by Blakie
\textit{et al.} \cite{gardiner} and Zambelli \textit{et al.}
\cite{stringari} and has been realized experimentally in Refs.~\onlinecite{Stamper, Vogel, Ozeri, Steinhauer, Katz, Muniz}.
\begin{figure}[t]
  \centering%
  \begin{subfloat}%
\includegraphics[width=.4\textwidth,angle=0]{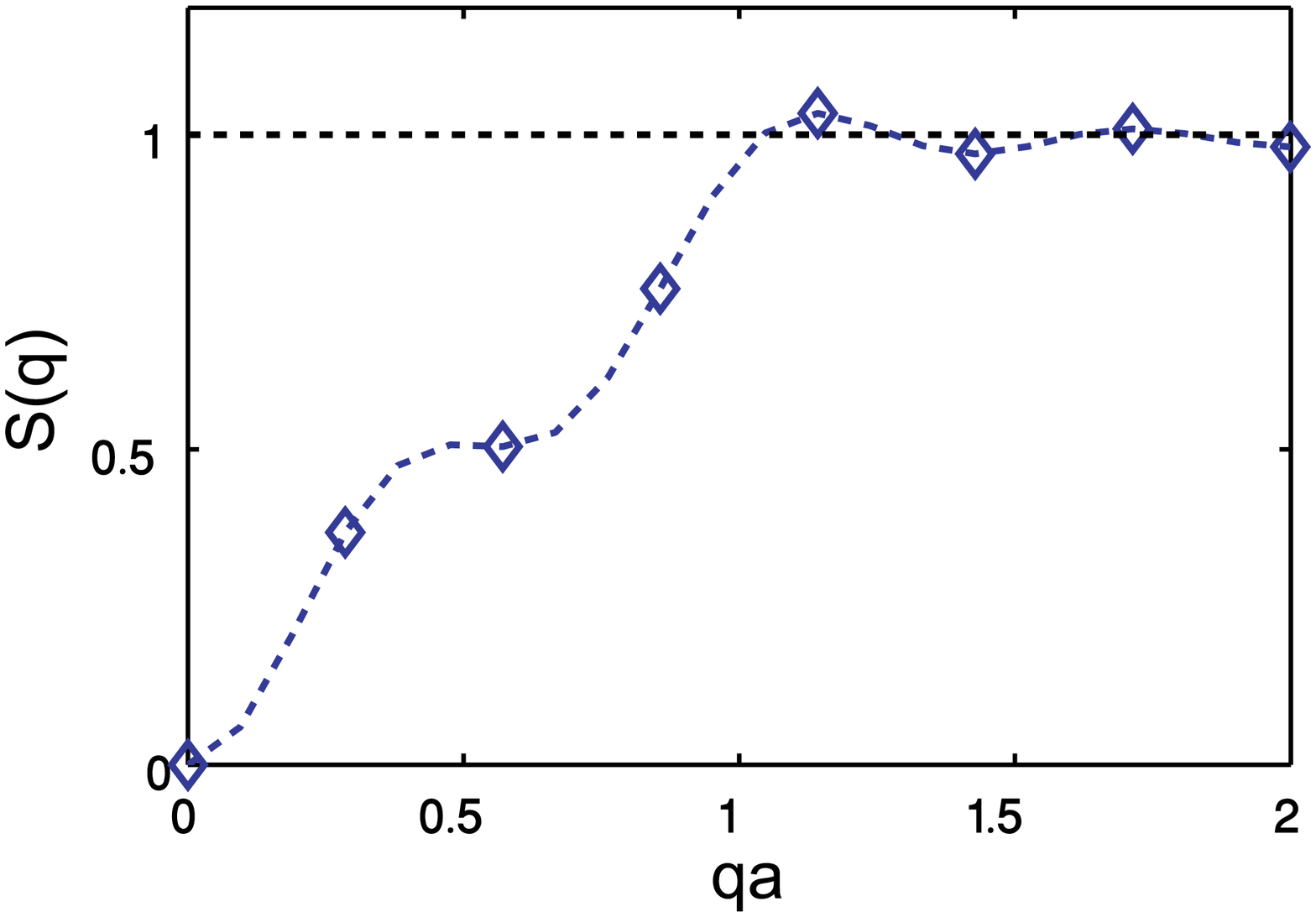}
\caption{$$}
 \end{subfloat}%
  \hspace{0cm}%
  \begin{subfloat}%

\includegraphics[width=.4\textwidth,angle=0]{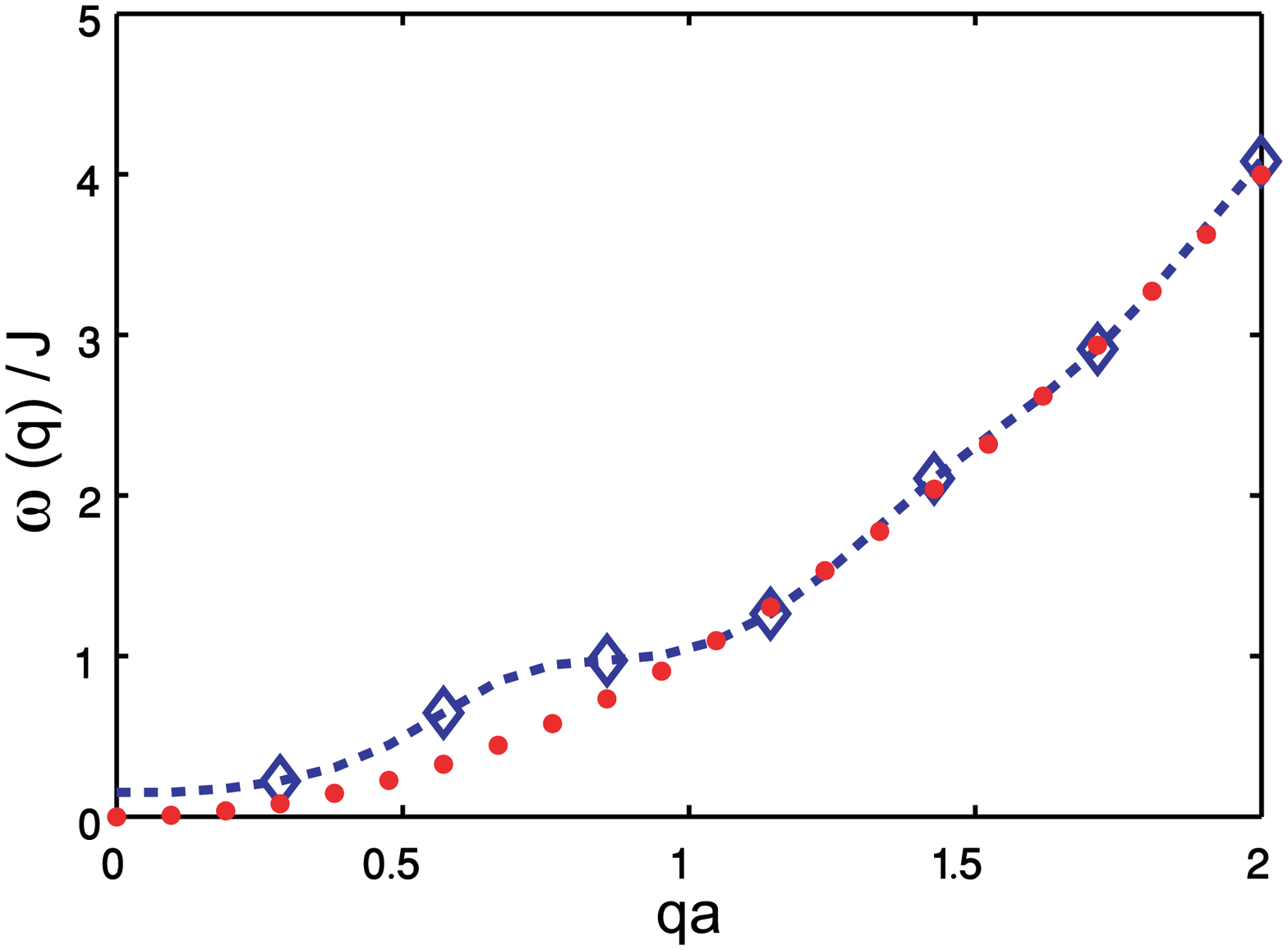}
\caption{$$}
   \end{subfloat}%
\caption{ (color online) (a) Structure factor and (b) energy spectrum for a 11x11
lattice with 3 atoms on a torus. Points shows the momentums allowed by the
boundary conditions. The
dotted line in (b) shows for comparison the low energy spectrum of a free particle which equals $J (qa)^2$. }
  \label{3atoms-structure}
\end{figure}

Also, Bragg spectroscopy in an optical lattice is discussed in
Ref.~\onlinecite{menotti} in a mean-field approach and also in
Ref.~\onlinecite{Rey} as a probe of the Mott-insulator excitation spectrum.
On the other hand, in the context of quantum Hall effect, the static structure factor has been studied for probing the magnetoroton excitations
\cite{girvin86} and charge density waves \cite{rezayiCDW}. The dynamic and static
structure factors are given respectively as

\<S(\vec{q},\omega)=\sum_{n ,0}| \langle n | \rho^\dagger(\vec{q})|
0\rangle |^2 \delta(\omega-E_n+E_0), \>

 \<S(\vec{q})=\sum_{n,0} |
\langle n | \rho^\dagger(\vec{q})| 0\rangle |^2 =\sum_{n,0} | \langle n
| \sum_{\vec{r_i}} e^{i \vec{q}\cdot\vec{r_i}}| 0\rangle |^2, \>
where the density fluctuation operator is defined as
$\rho^\dagger(\vec{q})=\sum_{m,n} \mathcal{A}_{\vec{q}}(m,n)
c^\dagger_m c_n $ and the coefficient are defined as Fourier
transforms of the  Wannier functions:
$\mathcal{A}_{\vec{q}}(m,n)=\int d^2\vec{r}~
e^{i\vec{q}\cdot\vec{r}} \phi^*(\vec{r}-\vec{r}_m)
\phi(\vec{r}-\vec{r}_n)$, where the Wannier function
$\phi(\vec{r}-\vec{r}_n)$ is the wave function of an atom localized
on a site centered at $\vec{r}_n$. Below, we focus on
 deep optical lattices, where
$\mathcal{A}_{\vec{q}}(m,n)=e^{i\vec{q}\cdot\vec{r}_m}
\delta_{m,n}$.

In the structure factor, there is a  sum over the excited states $|n\rangle$ and ground states $|0\rangle$ and the self-term is thus excluded.  The
ground state on a torus is two-fold degenerate and
therefore in our numerics, we  add the contribution of both.

Since we are working on a discrete lattice, there will be a
cut-off in the allowed momentum given by the lattice spacing $q_{\rm max}= \pi/a$, where ${\rm a}$ is the distance between lattice sites.
Fig.~\ \ref{3atoms-structure}(a) shows the structure factor for the
case $\nu=1/2$ for a small $\alpha$ calculated from our numerical
diagonalizations. In the data presented here, we have chosen
$\overrightarrow{q}=q \hat{x}$ and but
the result should be similar in
other directions in the lattice plane. We see that $S(q)$ is modulated at a
momentum corresponding to the magnetic length. For the parameters that we have investigated, the general features of the structure factor is independent of
the size of the system.

We obtain the excitation spectrum shown in Fig.\ \ref{3atoms-structure} (b) similar to Ref.~\onlinecite{girvin86} by
the Feynman-Bijl approximation. In the continuum limit ($\alpha \ll
1 $), we assume that the low-lying excitations are long-wavelength
density oscillations and their wave functions can be approximated to
have the form $ \propto \rho_k |0 \rangle$. Therefore, the
variational estimate for the excitation energy is $\omega(q) \simeq
\hbar^2q^2/~2 m S(q) $. At zero momentum and at the
momentum corresponding to the magnetic length, there are gaps, and
we also observe a deviation
from the free particle spectrum similar to the magneoroton case as a
reminiscent of the Wigner crystal. It should be noted that the
deviation does not depend on the size or the number of particles in
the system. As clearly seen in the Fig.~\ \ref{3atoms-structure} the
energy spectrum and structure factor
deviate from those of free particles, therefore, it could be used as an experimental probe of the system.

The structure factor and excitation spectrum imply some general
features that are very different from that of the Mott-insulator and
superfluid states, and can be used a powerful experimental indication of the
quantum Hall states.

\section{Generating Magnetic Hamiltonian for neutral atoms on a
lattice} \label{magnetic}

Recently, there has been several proposals for producing artificial
magnetic field for neutral atoms in
optical lattices \cite{jaksch, mueller,
sorensen}, however, the implementation of each of them is still
experimentally demanding.
Recently, there has been an experimental demonstration of a rotating optical lattice \cite{cornell_06} which is equivalent to an effective magnetic field (see below). The technique used in this experiment, however, generates a lattice with a large lattice spacing, because it uses laser beam which are not exactly counter propagating. This longer spacing reduces the energy scale in the lattice and thereby  also reduces quantities such as energy gaps. Here, we shall now introduce an alternative method for generating  a rotating optical lattice, which does not result in an increased lattice spacing. This method consists of rotating the optical
lattice by manipulating laser beams.

In a frame of reference rotating with angular velocity
$\omega$ around the $z-axis$, the Hamiltonian for a particle of mass $m$ in an (planar)
harmonic trap of natural frequency $\omega_0$ is

\begin{eqnarray} H&=&\frac{p^2}{2m}+\frac{1}{2} m \omega_0 (x^2+y^2)-\omega
\hat{z}.r\times p \nonumber\\
&=& \frac{(p-m\omega \hat{z}\times r
)^2}{2m}+\frac{1}{2}m(\omega_0^2-\omega^2)(x^2+y^2)
\end{eqnarray}
At resonance $\omega_0=\omega $ the form is equivalent to the
Hamiltonian of a particle of charge $q$ experiencing an effective
magnetic field $B=\bigtriangledown \times (m\omega \hat{z}\times
r/q)=(2m\omega/q) \hat{z}$.  Therefore, by simply rotating the
optical lattice, we can mimic the magnetic field for neutral atoms.

\begin{figure}

\includegraphics[width=.45\textwidth,angle=0]{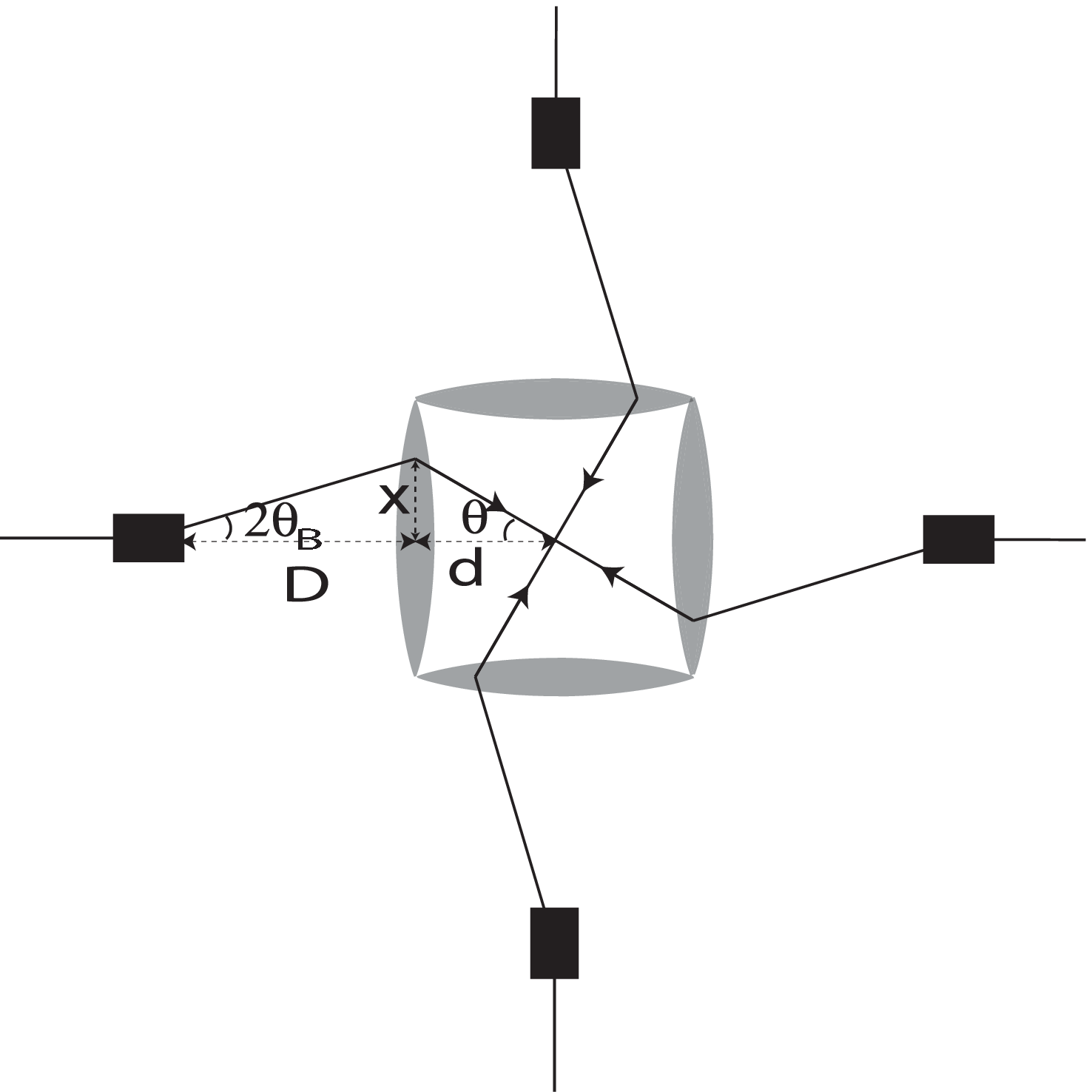}

 \caption{Proposal for realizing a rotating optical lattice. Four AOMs (black boxes) changes the direction of the lattice beam, which are subsequently focussed in the middle of the setup by four lenses (grey).
Simultaneously varying the four diffraction angles in the AOMs will
generate a rotating optical lattice.}

\label{lattice_rotation}
\end{figure}

To rotate the optical lattice, we propose to set up four acousto-optic
modulators (AOM) and four focusing composite lenses as shown in the Fig.~\ref{lattice_rotation}. By
sweeping the acoustic wave frequency, the beams can be focused and
make a rotating optical lattice.

In an AOM, for the first order diffracted light we have $ \sin
\theta_B=\frac{\lambda }{2 \Lambda}$, where $\Lambda$ is the wavelength of
sound in the medium, $\lambda$ is the light wavelength and
$\theta_B$ is half of  the angle  between a diffracted beam and the
non-diffracted beam (Fig.\ \ref{lattice_rotation}). By increasing
the frequency of the acoustic wave, the diffraction angle increases.
However, the beam should be focused by a large aperture lens so that
it always passes the region where we want to make the optical
lattice. By focusing a similar but counter-propagating beam, we can
make a rotating standing wave (Fig.\ \ref{lattice_rotation}). By
repeating the same configuration in the transverse direction, we can
make a rotating optical lattice. In particular, if the AOM is far
from the composite lenses, $D\gg d$, then $x/D=\lambda /  \Lambda$ and $x/d=\tan
\theta $ where $ -\pi/4 \leq (\theta=\omega t) \leq \pi/4$, where
the parameters are defined in Fig.\ \ref{lattice_rotation}. If we
consider a square lattice with dimensions $N_x=N_y=\mathcal{N}$, the
number of  magnetic flux given by rotation is \< N_{\phi}=\frac{B
A}{\Phi_0}= \frac{\pi}{2} \frac{\mathcal{N}^2 \omega}{\omega_r}\>
where $\omega_r=\hbar k^2/2M$ is the atomic recoil frequency.

On the other hand, the upper limit for the magnetic field comes from
the single Bloch band approximation which we made in writing the
Hamiltonian for the optical lattice. In order for the particles to
remain in the first band, the traveling lattice beams should move
them adiabatically. From Ref.\ \onlinecite{Salomon}, the
adiabaticity condition for a moving lattice with an acceleration
$\eta$ equal to $\omega^2 \mathcal{N} \lambda/4$ at the edge, is $ m
\eta \lambda \ll \hbar \frac{\omega_p^4}{\omega_r^3} $, where
$\omega_p$ is the frequency difference between the first and the
second band in the lattice. This puts a limit on how large the
lattice can become
 $ \mathcal{N} \ll \frac{\omega_p^4}{\omega^2 \omega_r^2}$.

Hence, for $\nu=1/2$  with a lattice filling fraction $
\frac{N}{\mathcal{N}^2} \sim \frac{1}{8}$ and a typical recoil
frequency $\omega_r=(2\pi) 4 $  kHz, one can enter the regime of
fractional quantum Hall effect by rotating the lattice at $\omega
\sim (2\pi) 650$ Hz. If a deep optical lattice is used e.g. $
\omega_p \sim 10 \omega_r $, the adiabaticity condition is easily
satisfied for a lattice of size $\mathcal{N} \sim1000$.  The
experimentally challenging part will, however, likely be to mitigate
the instability of the lattice caused by the thickness of the beam
and aberration of the lenses at the turning points i.e. near
$\theta=\pi/4$.

\section{Conclusions}

An extended study of the realization of the fractional quantum Hall
states in optical lattices has been presented. We showed that a
Hamiltonian similar to that of a magnetic field for charged
particles can be constructed for neutral ultra cold atoms and
molecules confined in an optical lattice. By adding an onsite
interaction for the case of $\nu=1/2$, an energy gap develops
between the ground state and the first excited state, which
increases linearly as $\alpha U$ and saturates to its value in the
hardcore limit $U \gg J$. We learned that the Laughlin wavefunction
is a reliable description of the system for low flux densities
$\alpha \lesssim 0.25$. However, for higher $\alpha$'s, the lattice
structure becomes more pronounced and a better description of the
system can be carried out by investigating the Chern number
associated to the ground state manifold. The Chern number indicates
that the system has topological order up to some critical flux
density $\alpha_c \simeq 0.4$, where the properties of the ground
state manifold starts to change. We have also studied $\nu=1/4$,
where compared to $\nu=1/2$, the Laughlin wavefunction only
describes the ground state for lower values of the flux $\alpha
\lesssim 0.1$. We showed that a dipole-dipole interaction can
enhance the gap and stabilize the system, and therefore make the
ground state more experimentally realizable. Bragg spectroscopy has
been studied as a potential experimental tool to diagnose theses
incompressible states.

Characterization of the ground state by evaluating the Chern number,
developed in Sec. \ref{gauge-fixing}, can be generalized to other
interesting many-body systems where the conventional overlap
calculation fails to work. In particular, this method can be applied
to ground states with non-Abelian statistics which are appealing
candidates for fault-tolerant quantum computation.

\begin{acknowledgments}

We thank K. Yang, M. Greiner, S. Girvin and J. I. Cirac for fruitful
discussions. This work was partially supported by the NSF Career
award, Packard Foundation, AFSOR and the Danish Natural Science
Research Council.
\end{acknowledgments}


\begin{thebibliography}{68}
\expandafter\ifx\csname natexlab\endcsname\relax\def\natexlab#1{#1}\fi
\expandafter\ifx\csname bibnamefont\endcsname\relax
  \def\bibnamefont#1{#1}\fi
\expandafter\ifx\csname bibfnamefont\endcsname\relax
  \def\bibfnamefont#1{#1}\fi
\expandafter\ifx\csname citenamefont\endcsname\relax
  \def\citenamefont#1{#1}\fi
\expandafter\ifx\csname url\endcsname\relax
  \def\url#1{\texttt{#1}}\fi
\expandafter\ifx\csname urlprefix\endcsname\relax\def\urlprefix{URL }\fi
\providecommand{\bibinfo}[2]{#2}
\providecommand{\eprint}[2][]{\url{#2}}

\bibitem[{\citenamefont{Greiner et~al.}(2002)\citenamefont{Greiner, Mandel,
  Esslinger, Hansch, and Bloch}}]{greiner}
\bibinfo{author}{\bibfnamefont{M.}~\bibnamefont{Greiner}},
  \bibinfo{author}{\bibfnamefont{O.}~\bibnamefont{Mandel}},
  \bibinfo{author}{\bibfnamefont{T.}~\bibnamefont{Esslinger}},
  \bibinfo{author}{\bibfnamefont{T.~W.} \bibnamefont{Hansch}},
  \bibnamefont{and} \bibinfo{author}{\bibfnamefont{I.}~\bibnamefont{Bloch}},
  \bibinfo{journal}{Nature} \textbf{\bibinfo{volume}{415}}, \bibinfo{pages}{39}
  (\bibinfo{year}{2002}).

\bibitem[{\citenamefont{Mandel et~al.}(2003)\citenamefont{Mandel, Greiner,
  Widera, Rom, Hansch, and Bloch}}]{Mandel03}
\bibinfo{author}{\bibfnamefont{O.}~\bibnamefont{Mandel}},
  \bibinfo{author}{\bibfnamefont{M.}~\bibnamefont{Greiner}},
  \bibinfo{author}{\bibfnamefont{A.}~\bibnamefont{Widera}},
  \bibinfo{author}{\bibfnamefont{T.}~\bibnamefont{Rom}},
  \bibinfo{author}{\bibfnamefont{T.~W.} \bibnamefont{Hansch}},
  \bibnamefont{and} \bibinfo{author}{\bibfnamefont{I.}~\bibnamefont{Bloch}},
  \bibinfo{journal}{Nature} \textbf{\bibinfo{volume}{425}},
  \bibinfo{pages}{937} (\bibinfo{year}{2003}).

\bibitem[{\citenamefont{Campbell et~al.}(2006)}]{Campbell06}
\bibinfo{author}{\bibfnamefont{G.~K.} \bibnamefont{Campbell}}
  \bibnamefont{et~al.}, \bibinfo{journal}{Phys. Rev. Lett.}
  \textbf{\bibinfo{volume}{96}}, \bibinfo{pages}{020406}
  (\bibinfo{year}{2006}).

\bibitem[{\citenamefont{Winkler et~al.}(2006)}]{Winkler06}
\bibinfo{author}{\bibfnamefont{K.}~\bibnamefont{Winkler}} \bibnamefont{et~al.},
  \bibinfo{journal}{Nature} \textbf{\bibinfo{volume}{441}},
  \bibinfo{pages}{853} (\bibinfo{year}{2006}).

\bibitem[{\citenamefont{Folling et~al.}(2005)}]{Folling05}
\bibinfo{author}{\bibfnamefont{S.}~\bibnamefont{Folling}} \bibnamefont{et~al.},
  \bibinfo{journal}{Nature} \textbf{\bibinfo{volume}{434}},
  \bibinfo{pages}{481} (\bibinfo{year}{2005}).

\bibitem[{\citenamefont{Abo-Shaeer et~al.}(2001)}]{aboshaeer}
\bibinfo{author}{\bibfnamefont{J.~R.} \bibnamefont{Abo-Shaeer}}
  \bibnamefont{et~al.}, \bibinfo{journal}{Science}
  \textbf{\bibinfo{volume}{292}}, \bibinfo{pages}{476} (\bibinfo{year}{2001}).

\bibitem[{\citenamefont{Wilkin and Gunn}(2000)}]{wilkin2000}
\bibinfo{author}{\bibfnamefont{N.~K.} \bibnamefont{Wilkin}} \bibnamefont{and}
  \bibinfo{author}{\bibfnamefont{J.}~\bibnamefont{Gunn}},
  \bibinfo{journal}{Phys. Rev. Lett.} \textbf{\bibinfo{volume}{84}},
  \bibinfo{pages}{6} (\bibinfo{year}{2000}).

\bibitem[{\citenamefont{Cooper et~al.}(2001)\citenamefont{Cooper, Wilkin, and
  Gunn}}]{cooper2001}
\bibinfo{author}{\bibfnamefont{N.~R.} \bibnamefont{Cooper}},
  \bibinfo{author}{\bibfnamefont{N.~K.} \bibnamefont{Wilkin}},
  \bibnamefont{and} \bibinfo{author}{\bibfnamefont{J.~M.~F.}
  \bibnamefont{Gunn}}, \bibinfo{journal}{Phys. Rev. Lett.}
  \textbf{\bibinfo{volume}{87}}, \bibinfo{pages}{120405}
  (\bibinfo{year}{2001}).

\bibitem[{\citenamefont{Rezayi et~al.}(2005)\citenamefont{Rezayi, Read, and
  Cooper}}]{cooper2005}
\bibinfo{author}{\bibfnamefont{E.~H.} \bibnamefont{Rezayi}},
  \bibinfo{author}{\bibfnamefont{N.}~\bibnamefont{Read}}, \bibnamefont{and}
  \bibinfo{author}{\bibfnamefont{N.~R.} \bibnamefont{Cooper}},
  \bibinfo{journal}{Phys. Rev. Lett.} \textbf{\bibinfo{volume}{95}},
  \bibinfo{pages}{160404} (\bibinfo{year}{2005}).

\bibitem[{\citenamefont{S{\o}rensen et~al.}(2005)\citenamefont{S{\o}rensen,
  Demler, and Lukin}}]{sorensen}
\bibinfo{author}{\bibfnamefont{A.}~\bibnamefont{S{\o}rensen}},
  \bibinfo{author}{\bibfnamefont{E.}~\bibnamefont{Demler}}, \bibnamefont{and}
  \bibinfo{author}{\bibfnamefont{M.}~\bibnamefont{Lukin}},
  \bibinfo{journal}{Phys.\ Rev.\ Lett.} \textbf{\bibinfo{volume}{94}},
  \bibinfo{pages}{086803} (\bibinfo{year}{2005}).

\bibitem[{\citenamefont{Kitaev}(2003)}]{kitaev}
\bibinfo{author}{\bibfnamefont{A.~Y.} \bibnamefont{Kitaev}},
  \bibinfo{journal}{Ann. Phys. (N.Y.)} \textbf{\bibinfo{volume}{303}},
  \bibinfo{pages}{2} (\bibinfo{year}{2003}).

\bibitem[{\citenamefont{Camino et~al.}(2005{\natexlab{a}})\citenamefont{Camino,
  Zhou, and Goldman}}]{goldman1}
\bibinfo{author}{\bibfnamefont{F.~E.} \bibnamefont{Camino}},
  \bibinfo{author}{\bibfnamefont{W.}~\bibnamefont{Zhou}}, \bibnamefont{and}
  \bibinfo{author}{\bibfnamefont{V.~J.} \bibnamefont{Goldman}},
  \bibinfo{journal}{Phys. Rev. Lett.} \textbf{\bibinfo{volume}{95}},
  \bibinfo{pages}{246802} (\bibinfo{year}{2005}{\natexlab{a}}).

\bibitem[{\citenamefont{Camino et~al.}(2005{\natexlab{b}})\citenamefont{Camino,
  Zhou, and Goldman}}]{goldman2}
\bibinfo{author}{\bibfnamefont{F.~E.} \bibnamefont{Camino}},
  \bibinfo{author}{\bibfnamefont{W.}~\bibnamefont{Zhou}}, \bibnamefont{and}
  \bibinfo{author}{\bibfnamefont{V.~J.} \bibnamefont{Goldman}},
  \bibinfo{journal}{Phys. Rev. B} \textbf{\bibinfo{volume}{74}},
  \bibinfo{pages}{115301} (\bibinfo{year}{2005}{\natexlab{b}}).

\bibitem[{\citenamefont{Paredes et~al.}(2001)\citenamefont{Paredes, Fedichev,
  Cirac, and Zoller}}]{paredes01}
\bibinfo{author}{\bibfnamefont{B.}~\bibnamefont{Paredes}},
  \bibinfo{author}{\bibfnamefont{P.}~\bibnamefont{Fedichev}},
  \bibinfo{author}{\bibfnamefont{J.~I.} \bibnamefont{Cirac}}, \bibnamefont{and}
  \bibinfo{author}{\bibfnamefont{P.}~\bibnamefont{Zoller}},
  \bibinfo{journal}{Phys. Rev. Lett.} \textbf{\bibinfo{volume}{87}},
  \bibinfo{pages}{010402} (\bibinfo{year}{2001}).

\bibitem[{\citenamefont{Hofstadter}(1976)}]{hofstadter}
\bibinfo{author}{\bibfnamefont{D.}~\bibnamefont{Hofstadter}},
  \bibinfo{journal}{Phys. Rev. B} \textbf{\bibinfo{volume}{14}},
  \bibinfo{pages}{2239} (\bibinfo{year}{1976}).

\bibitem[{\citenamefont{Laughlin}(1983)}]{laughlin}
\bibinfo{author}{\bibfnamefont{R.}~\bibnamefont{Laughlin}},
  \bibinfo{journal}{Phys. Rev. Lett.} \textbf{\bibinfo{volume}{50}},
  \bibinfo{pages}{1395} (\bibinfo{year}{1983}).

\bibitem[{\citenamefont{Jaksch et~al.}(1998)\citenamefont{Jaksch, Bruder,
  Cirac, Gardiner, and Zoller}}]{jaksch98}
\bibinfo{author}{\bibfnamefont{D.}~\bibnamefont{Jaksch}},
  \bibinfo{author}{\bibfnamefont{C.}~\bibnamefont{Bruder}},
  \bibinfo{author}{\bibfnamefont{J.~I.} \bibnamefont{Cirac}},
  \bibinfo{author}{\bibfnamefont{C.~W.} \bibnamefont{Gardiner}},
  \bibnamefont{and} \bibinfo{author}{\bibfnamefont{P.}~\bibnamefont{Zoller}},
  \bibinfo{journal}{Phys. Rev. Lett.} \textbf{\bibinfo{volume}{81}},
  \bibinfo{pages}{3108} (\bibinfo{year}{1998}).

\bibitem[{\citenamefont{Peierls}(1933)}]{peierls}
\bibinfo{author}{\bibfnamefont{R.~E.} \bibnamefont{Peierls}},
  \bibinfo{journal}{Z. Phys. Rev.} \textbf{\bibinfo{volume}{80}},
  \bibinfo{pages}{763} (\bibinfo{year}{1933}).

\bibitem[{\citenamefont{Bretin et~al.}(2004)\citenamefont{Bretin, Stock,
  Seurin, and Dalibard}}]{dalibard_04}
\bibinfo{author}{\bibfnamefont{V.}~\bibnamefont{Bretin}},
  \bibinfo{author}{\bibfnamefont{S.}~\bibnamefont{Stock}},
  \bibinfo{author}{\bibfnamefont{Y.}~\bibnamefont{Seurin}}, \bibnamefont{and}
  \bibinfo{author}{\bibfnamefont{J.}~\bibnamefont{Dalibard}},
  \bibinfo{journal}{Phys. Rev. Lett.} \textbf{\bibinfo{volume}{92}},
  \bibinfo{pages}{050403} (\bibinfo{year}{2004}).

\bibitem[{\citenamefont{Schweikhard et~al.}(2004)\citenamefont{Schweikhard,
  Coddington, Engels, Mogendorff, and Cornell}}]{cornell_04}
\bibinfo{author}{\bibfnamefont{V.}~\bibnamefont{Schweikhard}},
  \bibinfo{author}{\bibfnamefont{I.}~\bibnamefont{Coddington}},
  \bibinfo{author}{\bibfnamefont{P.}~\bibnamefont{Engels}},
  \bibinfo{author}{\bibfnamefont{V.~P.} \bibnamefont{Mogendorff}},
  \bibnamefont{and} \bibinfo{author}{\bibfnamefont{E.~A.}
  \bibnamefont{Cornell}}, \bibinfo{journal}{Phys. Rev. Lett.}
  \textbf{\bibinfo{volume}{92}}, \bibinfo{pages}{040404}
  (\bibinfo{year}{2004}).

\bibitem[{\citenamefont{Jaksch and Zoller}(2003)}]{jaksch}
\bibinfo{author}{\bibfnamefont{D.}~\bibnamefont{Jaksch}} \bibnamefont{and}
  \bibinfo{author}{\bibfnamefont{P.}~\bibnamefont{Zoller}},
  \bibinfo{journal}{New\ J.\ Phys.} \textbf{\bibinfo{volume}{5}},
  \bibinfo{pages}{Art. No. 56} (\bibinfo{year}{2003}).

\bibitem[{\citenamefont{Mueller}(2004)}]{mueller}
\bibinfo{author}{\bibfnamefont{E.~J.} \bibnamefont{Mueller}},
  \bibinfo{journal}{Phys.\ Rev.\ A} \textbf{\bibinfo{volume}{70}},
  \bibinfo{pages}{041603} (\bibinfo{year}{2004}).

\bibitem[{\citenamefont{Tung et~al.}(2006)\citenamefont{Tung, Schweikhard, and
  Cornell}}]{cornell_06}
\bibinfo{author}{\bibfnamefont{S.}~\bibnamefont{Tung}},
  \bibinfo{author}{\bibfnamefont{V.}~\bibnamefont{Schweikhard}},
  \bibnamefont{and} \bibinfo{author}{\bibfnamefont{E.~A.}
  \bibnamefont{Cornell}}, \bibinfo{journal}{Phys. Rev. Lett.}
  \textbf{\bibinfo{volume}{97}}, \bibinfo{pages}{240402}
  (\bibinfo{year}{2006}).

\bibitem[{\citenamefont{Popp et~al.}(2004)\citenamefont{Popp, Paredes, and
  Cirac}}]{paredes04}
\bibinfo{author}{\bibfnamefont{M.}~\bibnamefont{Popp}},
  \bibinfo{author}{\bibfnamefont{B.}~\bibnamefont{Paredes}}, \bibnamefont{and}
  \bibinfo{author}{\bibfnamefont{J.~I.} \bibnamefont{Cirac}},
  \bibinfo{journal}{Phys. Rev. A} \textbf{\bibinfo{volume}{70}},
  \bibinfo{pages}{053612} (\bibinfo{year}{2004}).

\bibitem[{\citenamefont{Donley et~al.}(2002)\citenamefont{Donley, Claussen,
  Thompson, and Wieman}}]{Donley02}
\bibinfo{author}{\bibfnamefont{E.~A.} \bibnamefont{Donley}},
  \bibinfo{author}{\bibfnamefont{N.~R.} \bibnamefont{Claussen}},
  \bibinfo{author}{\bibfnamefont{S.~T.} \bibnamefont{Thompson}},
  \bibnamefont{and} \bibinfo{author}{\bibfnamefont{C.~E.}
  \bibnamefont{Wieman}}, \bibinfo{journal}{Nature}
  \textbf{\bibinfo{volume}{417}}, \bibinfo{pages}{529} (\bibinfo{year}{2002}).

\bibitem[{\citenamefont{D\"urr et~al.}(2004)\citenamefont{D\"urr, Volz, Marte,
  and Rempe}}]{Durr04}
\bibinfo{author}{\bibfnamefont{S.}~\bibnamefont{D\"urr}},
  \bibinfo{author}{\bibfnamefont{T.}~\bibnamefont{Volz}},
  \bibinfo{author}{\bibfnamefont{A.}~\bibnamefont{Marte}}, \bibnamefont{and}
  \bibinfo{author}{\bibfnamefont{G.}~\bibnamefont{Rempe}},
  \bibinfo{journal}{Phys. Rev. Lett.} \textbf{\bibinfo{volume}{92}},
  \bibinfo{pages}{020406} (\bibinfo{year}{2004}).

\bibitem[{\citenamefont{Read and Rezayi}(1996)}]{read96}
\bibinfo{author}{\bibnamefont{Read}} \bibnamefont{and}
  \bibinfo{author}{\bibnamefont{Rezayi}}, \bibinfo{journal}{Phys. Rev. B}
  \textbf{\bibinfo{volume}{54}}, \bibinfo{pages}{16864} (\bibinfo{year}{1996}).

\bibitem[{\citenamefont{Haldane}(1985)}]{haldane85}
\bibinfo{author}{\bibfnamefont{F.}~\bibnamefont{Haldane}},
  \bibinfo{journal}{Phys. Rev. Lett.} \textbf{\bibinfo{volume}{55}},
  \bibinfo{pages}{2095} (\bibinfo{year}{1985}).

\bibitem[{\citenamefont{Haldane and Rezayi}(1985)}]{haldane85b}
\bibinfo{author}{\bibfnamefont{F.~D.~M.} \bibnamefont{Haldane}}
  \bibnamefont{and} \bibinfo{author}{\bibfnamefont{E.~H.}
  \bibnamefont{Rezayi}}, \bibinfo{journal}{Phys. Rev. Lett.}
  \textbf{\bibinfo{volume}{54}}, \bibinfo{pages}{237} (\bibinfo{year}{1985}).

\bibitem[{\citenamefont{Fradkin}(1991)}]{fradkin}
\bibinfo{author}{\bibfnamefont{E.}~\bibnamefont{Fradkin}},
  \emph{\bibinfo{title}{Field Theories of Condensed Matter System}}
  (\bibinfo{publisher}{Addison Wesley}, \bibinfo{year}{1991}).

\bibitem[{\citenamefont{Gerton et~al.}(2000)\citenamefont{Gerton, Strekalov,
  Prodan, and Hulet}}]{hulet}
\bibinfo{author}{\bibfnamefont{J.}~\bibnamefont{Gerton}},
  \bibinfo{author}{\bibfnamefont{M.}~\bibnamefont{Strekalov}},
  \bibinfo{author}{\bibfnamefont{I.}~\bibnamefont{Prodan}}, \bibnamefont{and}
  \bibinfo{author}{\bibfnamefont{R.}~\bibnamefont{Hulet}},
  \bibinfo{journal}{Nature} \textbf{\bibinfo{volume}{408}},
  \bibinfo{pages}{692} (\bibinfo{year}{2000}).

\bibitem[{\citenamefont{Avron et~al.}(1983)\citenamefont{Avron, Seiler, and
  Simon}}]{avron}
\bibinfo{author}{\bibfnamefont{J.}~\bibnamefont{Avron}},
  \bibinfo{author}{\bibfnamefont{R.}~\bibnamefont{Seiler}}, \bibnamefont{and}
  \bibinfo{author}{\bibfnamefont{B.}~\bibnamefont{Simon}},
  \bibinfo{journal}{Phys. Rev. Lett} \textbf{\bibinfo{volume}{51}},
  \bibinfo{pages}{51} (\bibinfo{year}{1983}).

\bibitem[{\citenamefont{Thouless et~al.}(1982)\citenamefont{Thouless, Kohmoto,
  Nightingale, and den Nijs}}]{TKNdN}
\bibinfo{author}{\bibfnamefont{D.~J.} \bibnamefont{Thouless}},
  \bibinfo{author}{\bibfnamefont{M.}~\bibnamefont{Kohmoto}},
  \bibinfo{author}{\bibfnamefont{M.~P.} \bibnamefont{Nightingale}},
  \bibnamefont{and} \bibinfo{author}{\bibfnamefont{M.}~\bibnamefont{den Nijs}},
  \bibinfo{journal}{Phys. Rev. Lett} \textbf{\bibinfo{volume}{49}},
  \bibinfo{pages}{405} (\bibinfo{year}{1982}).

\bibitem[{\citenamefont{Niu et~al.}(1985)\citenamefont{Niu, Thouless, and
  Wu}}]{niu85}
\bibinfo{author}{\bibfnamefont{Q.}~\bibnamefont{Niu}},
  \bibinfo{author}{\bibfnamefont{D.~J.} \bibnamefont{Thouless}},
  \bibnamefont{and} \bibinfo{author}{\bibfnamefont{Y.~S.} \bibnamefont{Wu}},
  \bibinfo{journal}{Phys. Rev. B} \textbf{\bibinfo{volume}{31}},
  \bibinfo{pages}{3372} (\bibinfo{year}{1985}).

\bibitem[{\citenamefont{Tao and Haldane}(1986)}]{tao}
\bibinfo{author}{\bibfnamefont{R.}~\bibnamefont{Tao}} \bibnamefont{and}
  \bibinfo{author}{\bibfnamefont{F.~D.~M.} \bibnamefont{Haldane}},
  \bibinfo{journal}{Phys. Rev. B} \textbf{\bibinfo{volume}{33}},
  \bibinfo{pages}{3844} (\bibinfo{year}{1986}).

\bibitem[{\citenamefont{Sheng et~al.}(2003)\citenamefont{Sheng, Wan, Rezayi,
  Yang, Bhatt, and Haldane}}]{Sheng03}
\bibinfo{author}{\bibfnamefont{D.~N.} \bibnamefont{Sheng}},
  \bibinfo{author}{\bibfnamefont{X.}~\bibnamefont{Wan}},
  \bibinfo{author}{\bibfnamefont{E.~H.} \bibnamefont{Rezayi}},
  \bibinfo{author}{\bibfnamefont{K.}~\bibnamefont{Yang}},
  \bibinfo{author}{\bibfnamefont{R.~N.} \bibnamefont{Bhatt}}, \bibnamefont{and}
  \bibinfo{author}{\bibfnamefont{F.~D.~M.} \bibnamefont{Haldane}},
  \bibinfo{journal}{Phys. Rev. Lett.} \textbf{\bibinfo{volume}{90}},
  \bibinfo{pages}{256802} (\bibinfo{year}{2003}).

\bibitem[{\citenamefont{Wen and Niu}(1990)}]{wen90}
\bibinfo{author}{\bibfnamefont{X.~G.} \bibnamefont{Wen}} \bibnamefont{and}
  \bibinfo{author}{\bibfnamefont{Q.}~\bibnamefont{Niu}},
  \bibinfo{journal}{Phys. Rev. B} \textbf{\bibinfo{volume}{41}},
  \bibinfo{pages}{9377} (\bibinfo{year}{1990}).

\bibitem[{\citenamefont{Wen}(1990)}]{wen90b}
\bibinfo{author}{\bibfnamefont{X.~G.} \bibnamefont{Wen}},
  \bibinfo{journal}{Phys. Rev. B} \textbf{\bibinfo{volume}{40}},
  \bibinfo{pages}{7387} (\bibinfo{year}{1990}).

\bibitem[{\citenamefont{Oshikawa and Senthil}(2006)}]{senthil}
\bibinfo{author}{\bibfnamefont{M.}~\bibnamefont{Oshikawa}} \bibnamefont{and}
  \bibinfo{author}{\bibfnamefont{T.}~\bibnamefont{Senthil}},
  \bibinfo{journal}{Phys. Rev. Lett.} \textbf{\bibinfo{volume}{96}},
  \bibinfo{pages}{060601} (\bibinfo{year}{2006}).

\bibitem[{\citenamefont{Kitaev}(2005)}]{kitaev2005}
\bibinfo{author}{\bibfnamefont{A.}~\bibnamefont{Kitaev}},
  \bibinfo{journal}{Ann. Phys. (N.Y.)} \textbf{\bibinfo{volume}{321}},
  \bibinfo{pages}{2} (\bibinfo{year}{2005}).

\bibitem[{\citenamefont{J.Bellissard et~al.}(1994)\citenamefont{J.Bellissard,
  van Elst, and Schulz-Baldes}}]{bellissard}
\bibinfo{author}{\bibnamefont{J.Bellissard}},
  \bibinfo{author}{\bibfnamefont{A.}~\bibnamefont{van Elst}}, \bibnamefont{and}
  \bibinfo{author}{\bibfnamefont{H.}~\bibnamefont{Schulz-Baldes}},
  \bibinfo{journal}{J. Math. Phys.} \textbf{\bibinfo{volume}{35}},
  \bibinfo{pages}{5373} (\bibinfo{year}{1994}).

\bibitem[{\citenamefont{Varnhagen}(1995)}]{varnhagen}
\bibinfo{author}{\bibfnamefont{R.}~\bibnamefont{Varnhagen}},
  \bibinfo{journal}{Nucl. Phys. B} \textbf{\bibinfo{volume}{443}},
  \bibinfo{pages}{501} (\bibinfo{year}{1995}).

\bibitem[{\citenamefont{Hatsugai}(2005)}]{hatsugai05}
\bibinfo{author}{\bibfnamefont{Y.}~\bibnamefont{Hatsugai}},
  \bibinfo{journal}{J. Phys. Jpn.} \textbf{\bibinfo{volume}{74}},
  \bibinfo{pages}{1374} (\bibinfo{year}{2005}).

\bibitem[{\citenamefont{Rezayi and Haldane}(1985)}]{impurity1}
\bibinfo{author}{\bibfnamefont{E.~H.} \bibnamefont{Rezayi}} \bibnamefont{and}
  \bibinfo{author}{\bibfnamefont{F.~D.~M.} \bibnamefont{Haldane}},
  \bibinfo{journal}{Phys. Rev. B} \textbf{\bibinfo{volume}{32}},
  \bibinfo{pages}{6924} (\bibinfo{year}{1985}).

\bibitem[{\citenamefont{Zhang et~al.}(1985)\citenamefont{Zhang, Vulovic, Guo,
  and Das~Sarma}}]{impurity2}
\bibinfo{author}{\bibfnamefont{F.~C.} \bibnamefont{Zhang}},
  \bibinfo{author}{\bibfnamefont{V.~Z.} \bibnamefont{Vulovic}},
  \bibinfo{author}{\bibfnamefont{Y.}~\bibnamefont{Guo}}, \bibnamefont{and}
  \bibinfo{author}{\bibfnamefont{S.}~\bibnamefont{Das~Sarma}},
  \bibinfo{journal}{Phys. Rev. B} \textbf{\bibinfo{volume}{32}},
  \bibinfo{pages}{6920} (\bibinfo{year}{1985}).

\bibitem[{\citenamefont{Kohmoto}(1985)}]{kohmoto}
\bibinfo{author}{\bibfnamefont{M.}~\bibnamefont{Kohmoto}},
  \bibinfo{journal}{Ann. Phys. (N.Y.)} \textbf{\bibinfo{volume}{160}},
  \bibinfo{pages}{343} (\bibinfo{year}{1985}).

\bibitem[{\citenamefont{Kol and Read}(1993)}]{Kol}
\bibinfo{author}{\bibfnamefont{A.}~\bibnamefont{Kol}} \bibnamefont{and}
  \bibinfo{author}{\bibfnamefont{N.}~\bibnamefont{Read}},
  \bibinfo{journal}{Phys. Rev. B} \textbf{\bibinfo{volume}{48}},
  \bibinfo{pages}{8890} (\bibinfo{year}{1993}).

\bibitem[{\citenamefont{Read and Rezayi}(1999)}]{read_rezayi99}
\bibinfo{author}{\bibfnamefont{N.}~\bibnamefont{Read}} \bibnamefont{and}
  \bibinfo{author}{\bibfnamefont{E.}~\bibnamefont{Rezayi}},
  \bibinfo{journal}{Phys. Rev. B} \textbf{\bibinfo{volume}{59}},
  \bibinfo{pages}{8084} (\bibinfo{year}{1999}).

\bibitem[{\citenamefont{Baranov et~al.}(2005)\citenamefont{Baranov, Osterloh,
  and Lewenstein}}]{baranov}
\bibinfo{author}{\bibfnamefont{M.~A.} \bibnamefont{Baranov}},
  \bibinfo{author}{\bibfnamefont{K.}~\bibnamefont{Osterloh}}, \bibnamefont{and}
  \bibinfo{author}{\bibfnamefont{M.}~\bibnamefont{Lewenstein}},
  \bibinfo{journal}{Phys. Rev. Lett.} \textbf{\bibinfo{volume}{94}},
  \bibinfo{pages}{070404} (\bibinfo{year}{2005}).

\bibitem[{\citenamefont{Giovanazzi et~al.}(2002)\citenamefont{Giovanazzi,
  G\"orlitz, and Pfau}}]{giovanazzi}
\bibinfo{author}{\bibfnamefont{S.}~\bibnamefont{Giovanazzi}},
  \bibinfo{author}{\bibfnamefont{A.}~\bibnamefont{G\"orlitz}},
  \bibnamefont{and} \bibinfo{author}{\bibfnamefont{T.}~\bibnamefont{Pfau}},
  \bibinfo{journal}{Phys. Rev. Lett.} \textbf{\bibinfo{volume}{89}},
  \bibinfo{pages}{130401} (\bibinfo{year}{2002}).

\bibitem[{\citenamefont{Griesmaier et~al.}(2005)\citenamefont{Griesmaier,
  Werner, Hensler, Stuhler, and Pfau}}]{griesmaier}
\bibinfo{author}{\bibfnamefont{A.}~\bibnamefont{Griesmaier}},
  \bibinfo{author}{\bibfnamefont{J.}~\bibnamefont{Werner}},
  \bibinfo{author}{\bibfnamefont{S.}~\bibnamefont{Hensler}},
  \bibinfo{author}{\bibfnamefont{J.}~\bibnamefont{Stuhler}}, \bibnamefont{and}
  \bibinfo{author}{\bibfnamefont{T.}~\bibnamefont{Pfau}},
  \bibinfo{journal}{Phys. Rev. Lett.} \textbf{\bibinfo{volume}{94}},
  \bibinfo{pages}{160401} (\bibinfo{year}{2005}).

\bibitem[{\citenamefont{Palmer and Jaksch}(2006)}]{Jaksch06}
\bibinfo{author}{\bibfnamefont{R.~N.} \bibnamefont{Palmer}} \bibnamefont{and}
  \bibinfo{author}{\bibfnamefont{D.}~\bibnamefont{Jaksch}},
  \bibinfo{journal}{Phys. Rev. Lett.} \textbf{\bibinfo{volume}{96}},
  \bibinfo{pages}{180407} (\bibinfo{year}{2006}).

\bibitem[{\citenamefont{Altman et~al.}(2004)\citenamefont{Altman, Demler, and
  Lukin}}]{Ehud}
\bibinfo{author}{\bibfnamefont{E.}~\bibnamefont{Altman}},
  \bibinfo{author}{\bibfnamefont{E.}~\bibnamefont{Demler}}, \bibnamefont{and}
  \bibinfo{author}{\bibfnamefont{M.~D.} \bibnamefont{Lukin}},
  \bibinfo{journal}{Phys. Rev. A} \textbf{\bibinfo{volume}{70}},
  \bibinfo{pages}{013603} (\bibinfo{year}{2004}).

\bibitem[{\citenamefont{Bhat et~al.}(2007)}]{bhat07}
\bibinfo{author}{\bibfnamefont{R.}~\bibnamefont{Bhat}} \bibnamefont{et~al.},
  \bibinfo{journal}{cond-mat/07053341}  (\bibinfo{year}{2007}).

\bibitem[{\citenamefont{Stenger et~al.}(1999)}]{ketterle}
\bibinfo{author}{\bibfnamefont{J.}~\bibnamefont{Stenger}} \bibnamefont{et~al.},
  \bibinfo{journal}{Phys. Rev. Lett.} \textbf{\bibinfo{volume}{82}},
  \bibinfo{pages}{4569} (\bibinfo{year}{1999}).

\bibitem[{\citenamefont{Blakie et~al.}(2000)\citenamefont{Blakie, Ballagh, and
  Gardiner}}]{gardiner}
\bibinfo{author}{\bibfnamefont{P.~B.} \bibnamefont{Blakie}},
  \bibinfo{author}{\bibfnamefont{R.~J.} \bibnamefont{Ballagh}},
  \bibnamefont{and} \bibinfo{author}{\bibfnamefont{C.~W.}
  \bibnamefont{Gardiner}}, \bibinfo{journal}{Phys. Rev. A}
  \textbf{\bibinfo{volume}{65}}, \bibinfo{pages}{033602}
  (\bibinfo{year}{2000}).

\bibitem[{\citenamefont{Zambelli et~al.}(2000)\citenamefont{Zambelli,
  Pitaevskii, Stamper-Kurn, and Stringari}}]{stringari}
\bibinfo{author}{\bibfnamefont{F.}~\bibnamefont{Zambelli}},
  \bibinfo{author}{\bibfnamefont{L.}~\bibnamefont{Pitaevskii}},
  \bibinfo{author}{\bibfnamefont{D.~M.} \bibnamefont{Stamper-Kurn}},
  \bibnamefont{and}
  \bibinfo{author}{\bibfnamefont{S.}~\bibnamefont{Stringari}},
  \bibinfo{journal}{Phys. Rev. A} \textbf{\bibinfo{volume}{61}},
  \bibinfo{pages}{063608} (\bibinfo{year}{2000}).

\bibitem[{\citenamefont{Stamper-Kurn et~al.}(1999)\citenamefont{Stamper-Kurn,
  Chikkatur, G\"orlitz, Inouye, Gupta, Pritchard, and Ketterle}}]{Stamper}
\bibinfo{author}{\bibfnamefont{D.~M.} \bibnamefont{Stamper-Kurn}},
  \bibinfo{author}{\bibfnamefont{A.~P.} \bibnamefont{Chikkatur}},
  \bibinfo{author}{\bibfnamefont{A.}~\bibnamefont{G\"orlitz}},
  \bibinfo{author}{\bibfnamefont{S.}~\bibnamefont{Inouye}},
  \bibinfo{author}{\bibfnamefont{S.}~\bibnamefont{Gupta}},
  \bibinfo{author}{\bibfnamefont{D.~E.} \bibnamefont{Pritchard}},
  \bibnamefont{and} \bibinfo{author}{\bibfnamefont{W.}~\bibnamefont{Ketterle}},
  \bibinfo{journal}{Phys. Rev. Lett.} \textbf{\bibinfo{volume}{83}},
  \bibinfo{pages}{2876} (\bibinfo{year}{1999}).

\bibitem[{\citenamefont{Vogels et~al.}(2002)\citenamefont{Vogels, Xu, Raman,
  Abo-Shaeer, and Ketterle}}]{Vogel}
\bibinfo{author}{\bibfnamefont{J.~M.} \bibnamefont{Vogels}},
  \bibinfo{author}{\bibfnamefont{K.}~\bibnamefont{Xu}},
  \bibinfo{author}{\bibfnamefont{C.}~\bibnamefont{Raman}},
  \bibinfo{author}{\bibfnamefont{J.~R.} \bibnamefont{Abo-Shaeer}},
  \bibnamefont{and} \bibinfo{author}{\bibfnamefont{W.}~\bibnamefont{Ketterle}},
  \bibinfo{journal}{Phys. Rev. Lett.} \textbf{\bibinfo{volume}{88}},
  \bibinfo{pages}{060402} (\bibinfo{year}{2002}).

\bibitem[{\citenamefont{Ozeri et~al.}(2002)\citenamefont{Ozeri, Steinhauer,
  Katz, and Davidson}}]{Ozeri}
\bibinfo{author}{\bibfnamefont{R.}~\bibnamefont{Ozeri}},
  \bibinfo{author}{\bibfnamefont{J.}~\bibnamefont{Steinhauer}},
  \bibinfo{author}{\bibfnamefont{N.}~\bibnamefont{Katz}}, \bibnamefont{and}
  \bibinfo{author}{\bibfnamefont{N.}~\bibnamefont{Davidson}},
  \bibinfo{journal}{Phys. Rev. Lett.} \textbf{\bibinfo{volume}{88}},
  \bibinfo{pages}{220401} (\bibinfo{year}{2002}).

\bibitem[{\citenamefont{Steinhauer et~al.}(2003)\citenamefont{Steinhauer, Katz,
  Ozeri, Davidson, Tozzo, and Dalfovo}}]{Steinhauer}
\bibinfo{author}{\bibfnamefont{J.}~\bibnamefont{Steinhauer}},
  \bibinfo{author}{\bibfnamefont{N.}~\bibnamefont{Katz}},
  \bibinfo{author}{\bibfnamefont{R.}~\bibnamefont{Ozeri}},
  \bibinfo{author}{\bibfnamefont{N.}~\bibnamefont{Davidson}},
  \bibinfo{author}{\bibfnamefont{C.}~\bibnamefont{Tozzo}}, \bibnamefont{and}
  \bibinfo{author}{\bibfnamefont{F.}~\bibnamefont{Dalfovo}},
  \bibinfo{journal}{Phys. Rev. Lett.} \textbf{\bibinfo{volume}{90}},
  \bibinfo{pages}{060404} (\bibinfo{year}{2003}).

\bibitem[{\citenamefont{Katz et~al.}(2004)\citenamefont{Katz, Ozeri,
  Steinhauer, Davidson, Tozzo, and Dalfovo}}]{Katz}
\bibinfo{author}{\bibfnamefont{N.}~\bibnamefont{Katz}},
  \bibinfo{author}{\bibfnamefont{R.}~\bibnamefont{Ozeri}},
  \bibinfo{author}{\bibfnamefont{J.}~\bibnamefont{Steinhauer}},
  \bibinfo{author}{\bibfnamefont{N.}~\bibnamefont{Davidson}},
  \bibinfo{author}{\bibfnamefont{C.}~\bibnamefont{Tozzo}}, \bibnamefont{and}
  \bibinfo{author}{\bibfnamefont{F.}~\bibnamefont{Dalfovo}},
  \bibinfo{journal}{Phys. Rev. lett.} \textbf{\bibinfo{volume}{93}},
  \bibinfo{pages}{220403} (\bibinfo{year}{2004}).

\bibitem[{\citenamefont{Muniz et~al.}(2006)\citenamefont{Muniz, Naik, and
  Raman}}]{Muniz}
\bibinfo{author}{\bibfnamefont{S.~R.} \bibnamefont{Muniz}},
  \bibinfo{author}{\bibfnamefont{D.~S.} \bibnamefont{Naik}}, \bibnamefont{and}
  \bibinfo{author}{\bibfnamefont{C.}~\bibnamefont{Raman}},
  \bibinfo{journal}{Phys. Rev. A} \textbf{\bibinfo{volume}{73}},
  \bibinfo{pages}{041605} (\bibinfo{year}{2006}).

\bibitem[{\citenamefont{Menotti et~al.}(2003)}]{menotti}
\bibinfo{author}{\bibfnamefont{C.}~\bibnamefont{Menotti}} \bibnamefont{et~al.},
  \bibinfo{journal}{Phys. Rev. A} \textbf{\bibinfo{volume}{67}},
  \bibinfo{pages}{053609} (\bibinfo{year}{2003}).

\bibitem[{\citenamefont{Rey et~al.}(2005)\citenamefont{Rey, Blakie, Pupillo,
  Williams, and Clark}}]{Rey}
\bibinfo{author}{\bibfnamefont{A.~M.} \bibnamefont{Rey}},
  \bibinfo{author}{\bibfnamefont{P.~B.} \bibnamefont{Blakie}},
  \bibinfo{author}{\bibfnamefont{G.}~\bibnamefont{Pupillo}},
  \bibinfo{author}{\bibfnamefont{C.~J.} \bibnamefont{Williams}},
  \bibnamefont{and} \bibinfo{author}{\bibfnamefont{C.~W.} \bibnamefont{Clark}},
  \bibinfo{journal}{Phys. Rev. A} \textbf{\bibinfo{volume}{72}},
  \bibinfo{pages}{023407} (\bibinfo{year}{2005}).

\bibitem[{\citenamefont{Girvin et~al.}(1986)}]{girvin86}
\bibinfo{author}{\bibfnamefont{M.~S.} \bibnamefont{Girvin}}
  \bibnamefont{et~al.}, \bibinfo{journal}{Phys. Rev. B}
  \textbf{\bibinfo{volume}{33}}, \bibinfo{pages}{2481} (\bibinfo{year}{1986}).

\bibitem[{\citenamefont{Rezayi et~al.}(1999)\citenamefont{Rezayi, Haldane, and
  Yang}}]{rezayiCDW}
\bibinfo{author}{\bibfnamefont{E.~H.} \bibnamefont{Rezayi}},
  \bibinfo{author}{\bibfnamefont{F.~D.~M.} \bibnamefont{Haldane}},
  \bibnamefont{and} \bibinfo{author}{\bibfnamefont{K.}~\bibnamefont{Yang}},
  \bibinfo{journal}{Phys. Rev. Lett.} \textbf{\bibinfo{volume}{83}},
  \bibinfo{pages}{1219} (\bibinfo{year}{1999}).

\bibitem[{\citenamefont{Ben~Dahan et~al.}(1996)\citenamefont{Ben~Dahan, Peik,
  Reichel, Castin, and Salomon}}]{Salomon}
\bibinfo{author}{\bibfnamefont{M.}~\bibnamefont{Ben~Dahan}},
  \bibinfo{author}{\bibfnamefont{E.}~\bibnamefont{Peik}},
  \bibinfo{author}{\bibfnamefont{J.}~\bibnamefont{Reichel}},
  \bibinfo{author}{\bibfnamefont{Y.}~\bibnamefont{Castin}}, \bibnamefont{and}
  \bibinfo{author}{\bibfnamefont{C.}~\bibnamefont{Salomon}},
  \bibinfo{journal}{Phys. Rev. Lett.} \textbf{\bibinfo{volume}{76}},
  \bibinfo{pages}{4508} (\bibinfo{year}{1996}).

\end{thebibliography}
\end{document}